\documentclass[aps,preprint,groupedaddress,floatfix]{revtex4-2}
\usepackage{amsmath,amsfonts,amsthm}
\usepackage{physics}
\usepackage[version=4]{mhchem}
\usepackage{graphicx}
\usepackage[T1]{fontenc}
\usepackage{tgbonum}
\usepackage[inline]{enumitem}
\begin{document}
%\preprint{}

\title{Quasiparticle electronic structure and optical response ($G_0W_0$+BSE) of anatase \ce{TiO2} starting from modified HSE06 functionals}


\author{Sruthil Lal S.B}
% \email[]{getsruthil@gmail.com}
%\thanks{}
\affiliation{Department of Physics, Pondicherry University, R. V. Nagar, Kalapet, Puducherry, India}

\author{D Murali}
% \email[]{getsruthil@gmail.com}
%\thanks{}
\affiliation{Indian Institute of Information Technology Design and Manufacturing (IIITDM), Kurnool, Andhra Pradesh, India}

\author{Matthias Posselt}
% \email[]{getsruthil@gmail.com}
%\thanks{}
\affiliation{Helmholtz-Zentrum Dresden-Rossendorf, Bautzner Landstraße 400, 01328 Dresden, Germany}

\author{Assa Aravindh Sasikala Devi}
% \email[]{getsruthil@gmail.com}
%\thanks{}
\affiliation{ Nano and molecular systems research unit, P.O.Box 8000, FI-90014, University of Oulu, Oulu, Finland}

\author{Alok Sharan}
\email[]{aloksharan@gmail.com}
%\thanks{}
\affiliation{Department of Physics, Pondicherry University, R. V. Nagar, Kalapet, Puducherry, India}


\date{\today}

\begin{abstract}
    The quasiparticle electronic structure and optical excitation of anatase \ce{TiO2} is determined within the framework of many-body perturbation theory (MBPT) 
	by combining the $G_0W_0$ method and the Bethe-Salpeter Equation (BSE). A modified version of the HSE06 screened hybrid functional, that includes 20\% exact 
	Fock exchange (HSE06(20)) as opposed to 25\% in the standard HSE06 functional, is used to set up the starting Hamiltonian for $G_0W_0$+BSE calculations. 
	The HSE06(20) functional accurately predicts the ground state electronic band structure. BSE calculations based on data from $G_0W_0$+HSE06(20) yield direct 
	optical excitation energies and oscillator strengths in excellent agreement with existing experiments and theoretical calculations characterizing direct 
	excitation. In particular, an exciton binding energy of 229 $\pm$ 10 meV is obtained, in close agreement with experiments. The projections of excitonic 
	states onto the quasiparticle band structure in a fatband representation shows that the lowest optical transition of anatase \ce{TiO2} consists of excitons 
	originating from the mixing of direct transitions within band pairs running parallel to the $\Gamma -Z $ direction in the tetragonal Brillouin zone. 
	This implies a strong spatial localization of excitons in the $xy$ plane of the lattice. This investigation highlights the importance of a suitable 
	non-interacting Hamiltonian for the MBPT based quasiparticle $G_0W_0$ and subsequent BSE calculations and suggests HSE06(20) as an optimal choice in the case of anatase \ce{TiO2}.  
\end{abstract}

%\keywords{}

\maketitle
\section{Introduction}
In the ever growing arena of renewable energy and photo-catalysis, Titanium Dioxide (\ce{TiO2}) is one of the most widely explored metal-oxides owing to its charge transport and oxidation properties along with its abundance, non-toxicity, and corrosion resistance. \ce{TiO2} finds promising applications in areas such as photo-generation of hydrogen from water\cite{fujishima1972electrochemical,fujishima2000titanium,mills2003preparation}, photo-active and opto-electronic devices\cite{bai2014titanium,chen2014electrospun}, dye-sensitized solar cells(DSSC)\cite{gratzel2005solar,bai2014titanium,o1991low,hagfeldt2000molecular}, degradation of pollutants under visible-light irradiation\cite{thompson2006surface,chen2007titanium}, production of hydrocarbon fuels\cite{khan2002efficient,varghese2009high}, antimicrobial coatings\cite{alotaibi2020enhanced}, nonlinear optical applications\cite{yaqub2021nonlinear} and so on. Among the naturally occurring \ce{TiO2} polymorphs, the anatase form is generally regarded as the more active phase in photo-catalysis than its other polymorphs-rutile and brookite\cite{xu2011photocatalytic,carp2004photoinduced,luttrell2014anatase,sumita2002photo}. Thin films of anatase has a wider optical-absorption gap and smaller electron effective mass that presumably leads to its higher charge carrier mobility\cite{gratzel2005solar,tang1994electrical}. Anatase \ce{TiO2} plays a key role in the injection process of photo-chemical solar cells with high conversion efficiency\cite{chen2014electrospun,bai2014titanium,gratzel2005solar,o1991low}. Tailor-made applications of these physico-chemical properties\cite{kang2019titanium} can be achieved by tuning the photoresponse by doping with noble gas\cite{pelaez2012review}, various metals\cite{alotaibi2020enhanced,atambo2019electronic,kulczyk2014investigation,li2016colossal,ma2021theoretical} as well as by strain engineering\cite{thulin2008calculations,maier2020flexoelectricity}. 

Rigorous approaches that goes beyond mean-field theory are necessary to quantitatively model photoemission and optical absorption in pristine anatase \ce{TiO2}\cite{atambo2019electronic,chiodo2010self,kang2010quasiparticle,landmann2012electronic,baldini2017strongly,tosoni2020structural}. The important physical quantities which are related to electrical and optical excitation in anatase \ce{TiO2} are accurately determined computationally within the many-body perturbation theory (MBPT)\cite{onida2002electronic,rohlfing2000electron,strinati1988application}. Within the MBPT framework, the $GW$ approximation (GWA)\cite{hedin1965new,aulbur2000quasiparticle,aryasetiawan1998gw,onida2002electronic,hybertsen1985first,godby1986accurate,hybertsen1986electron,shaltaf2008band,van2013gw,nguyen2012improving,umari2009optimal,umari2010gw} is commonly used for calculating quasiparticle energies and the Bethe-Salpeter equation (BSE) to include excitonic effects\cite{fuchs2007quasiparticle,salpeter1951relativistic,hanke1979many}. $GW$ method in combination with the Bethe–Salpeter equation (BSE)\cite{fuchs2007quasiparticle,salpeter1951relativistic} represents a standard many-body approach for the accurate determination of the band structure and optical properties of semiconductors \cite{onida2002electronic}. The simplest and  most efficient computational implementation of $GW$ is the single-shot-$GW$, often referred to as $G_0W_0$\cite{rostgaard2010fully,caruso2012unified,shishkin2007self,kotani2007quasiparticle,van2006quasiparticle,holm1998fully,tamme1999comment}. In $G_0W_0$ method, quasiparticle eigenvalues are estimated from a single $GW$ iteration, by perturbative Taylor expansion of self-energy around the DFT single-particle energies\cite{shishkin2006implementation,morales2017performance,hybertsen1985first,hybertsen1986electron,bechstedt1992efficient}. Although $G_0W_0$ is  proven to be  accurate enough for determining eigen energies, bandgaps, bandwidths, and band dispersion for semiconductors\cite{morales2017performance,hybertsen1985first,hybertsen1986electron,bechstedt1992efficient} the lack of self-consistency in $G_0W_0$ makes its extremely sensitive to the DFT initial wavefunction. Hence this choice has to be made carefully in order to capture the "right" amount of electronic screening\cite{korzdorfer2012strategy,bruneval2013benchmarking}. 

Compared to LDA/GGA\cite{sham1985density,perdew1981self,perdew1982density} or DFT+U\cite{hubbard1963electron}, DFT wavefunction with hybrid exchange and correlation(XC) functionals\cite{heyd2003hybrid,heyd2004assessment,krukau2006influence} are widely regarded as the most reasonable starting point for $G_0W_0$ calculations\cite{bechstedt2009ab,schleife2011tin,rodl2009quasiparticle,fuchs2008indium,fuchs2007quasiparticle,cappellini2013electronic,isseroff2012importance}. Hybrid functionals formally describe non-local two-particle scattering processes explicitly\cite{friedrich2012hybrid,furthmuller2005band},  and therefore, the resulting wavefunction are already close to the quasiparticle wavefunction. It remedies the bandgap underestimation by LDA/GGA functionals\cite{perdew2017understanding,sham1985density,perdew1981self,perdew1982density,perdew1983physical,heaton1982self, norman1983simplified,wang1983density,vydrov2005ionization,vydrov2006scaling,vydrov2004effect,seidl1996generalized,cohen2008fractional,ruzsinszky2006spurious,chan2010efficient}. With regards to \ce{TiO2} polymorphs, hybrid functionals do not cause lattice distortions\cite{anisimov1997first,gillen2013accurate}, which is otherwise one of the shortcomings of DFT+U when sub-optimal U values are used\cite{anisimov1997first,gillen2013accurate,raghav2020intrinsic}. As far as Ti-3\textit{d} states in anatase \ce{TiO2} and other 3\textit{d} transition-metal compounds are concerned, it is important that the starting wavefunction must describe its localized character and true nature of bonding\cite{mori2008localization,mattioli2010deep,unal2014electronic,na2006first,zhang2016all,aryasetiawan1995electronic,faleev2004all,bruneval2006exchange}. In comparison to LDA/GGA and DFT+U\cite{hubbard1963electron} methods, hybrid functionals reproduce the correct bandgap, localization of orbitals, ordering and occupation of bands leading to  much meaningful quasiparticle properties and dielectric functions\cite{bechstedt2009ab,fuchs2007quasiparticle,cappellini2013electronic,basera2019self,deak2018calculating}.

To date, optical properties of pristine anatase \ce{TiO2} taking the effect of many-body interactions into account have  been addressed only by few computational studies\cite{chiodo2010self,kang2010quasiparticle,landmann2012electronic,zhu2014stability,baldini2017strongly}. Majority of these studies rely on GGA(PBE) DFT wavefunction as starting point. To the best of our knowledge, apart from a recent study by \citeauthor[]{basera2019self}\cite{basera2019self}, there has been no theoretical study of the optical response of pristine anatase \ce{TiO2} exploring the role of hybrid functionals starting points in the MBPT calculations. In most of these studies, however, incongruous comparison of direct optical gap of \ce{TiO2} obtained by solving BSE is made with indirect optical gap, and binding energy of excitons are underestimated. 

We investigate the use of hybrid DFT wavefunction as starting point of MBPT calculations of quasiparticle excitation and excitonic optical spectra of pristine anatase \ce{TiO2}. HSE06 screened hybrid functional\cite{heyd2003hybrid,ge2006erratum} DFT Hamiltonian has been employed as the starting point for MBPT calculations. Single particle excitation are calculated by $G_0W_0$ and excitonic effects in the optical response of anatase \ce{TiO2} are treated within the BSE framework. Standard recipes of hybrid functionals prescribe the portion of exact exchange($\alpha$) to be fixed independent of the system being investigated. However, in general, $\alpha$ is material dependent\cite{alkauskas2008band,alkauskas2011defect,alkauskas2008defect,de2002effect,gerosa2015electronic} and is related to dielectric screening of the material\cite{marques2011density}. As we show in the present work, the optical response of anatase \ce{TiO2} is extremely tunable by adjusting the internal parameters of Hybrid functionals. In search of the best hybrid functional wavefunction for MBPT calculation, we optimize the fraction of exact exchange in the HSE06 functional in such a way that the modified functional correctly predicts both the ground state electronic structure and excited state response of pristine anatase \ce{TiO2}. The results of MBPT calculation with the proposed modified HSE06 functional is then discussed and compared with the existing literature. This study will act as a reliable benchmark of MBPT calculation for \ce{TiO2} and its possible extension at the same level of theory for all future calculations. Our investigation can benefit description of optical excitation of defective structures for which hybrid functionals are regularly employed \cite{chiodo2010self,basera2019self}.

%%%%%%%%%%%%% Organization of the article
The article is organized as follows. The general methodologies and computational schemes for DFT and MBPT calculations  are  summarized in Sec. \ref{sec:methods}. Sec. \ref{sec:results} presents results of optimization of DFT electronic structure (Sec. \ref{sec:optimizing_setup}), Quasiparticle correction (Sec. \ref{sec:QP_corrections}) and excitonic optical spectra (Sec. \ref{sec:optical_transitions}) computed with our proposed modified HSE06 hybrids as starting points. Key approximations used in the study and numerical details pertaining to the calculation has also been discussed wherever necessary. We conclude the article in Sec. \ref{sec:conclusion} with a short discussion of our results, along with its limitations including the neglect of role of coupling of electronic excitation to phonons.

\section{\label{sec:methods} Methodology and Computational details }

Density functional theory (DFT) calculations employing hybrid functional are  combined with methods based on many-body perturbation theory (MBPT). All ground state DFT, $GW$, and BSE calculations are performed using the Vienna ab-initio simulation package (VASP) \cite{kresse1993ab,hafner2008ab,kresse1996efficient,kresse1996efficiency}. Recommended PAW potentials\cite{blochl1994projector,kresse1999ultrasoft} supplied by VASP were employed for all atoms to describe the core-valence interactions. These PAW potentials are optimized for $GW$, as opposed to standard PAW potentials, and provide accurate description of scattering properties even at higher energies\cite{shishkin2006implementation}. Besides the Ti (4\textit{s} \& 3\textit{d}) and O (2\textit{s} \& 2\textit{p}) valence states, the shallow Ti (3\textit{s} and 3\textit{p}) core states are also treated as valence electrons. Their inclusion is essential for accurate calculation of quasiparticle energy gaps\cite{kang2010quasiparticle}, the omission of which is shown to increase the quasiparticle bandgap as high as 0.3 eV\cite{kang2015influence}. Gaussian smearing with $\sigma = 0.05$ eV was used to broaden the one electron levels. Based on the convergence of total free energy, we have set 520 eV as kinetic energy cut-off to describe the plane waves included in the basis set. 
An un-shifted $\Gamma$-centered $6\times 6 \times 3 $ Monkhorst-Pack grid \cite{monkhorst1976special} is used for sampling the body-centered tetragonal Brillouin zone of anatase \ce{TiO2}. Initially, geometry of the conventional unit cell of bulk anatase \ce{TiO2} is relaxed by using the GGA (PBE) functional. During geometry relaxation, the volume, cell shape, and ions were allowed to relax until the force and energy becomes less than $10^{-5} $ eV/{\AA} and $10^{-6} $ eV per atom, respectively. The relaxed tetragonal lattice is characterized by lattice constants: $a$ = 3.805 {\AA} and $c$ = 9.781 {\AA}, which are in reasonable agreement with lattice parameters measured in experiments\cite{burdett1987structural,rao1970thermal,howard1991structural}.

The electronic structure of optimized anatase \ce{TiO2} is then calculated by introducing non-local part of the exchange as in hybrid density functionals \cite{heyd2003hybrid,heyd2004assessment,krukau2006influence}. In this work HSE06 screened hybrid functional as parametrized by Heyd-Scuseria-Ernzerhof \cite{heyd2003hybrid,heyd2004efficient,ge2006erratum,heyd2006hybrid,paier2006screened,krukau2006influence} and its modifications are used to represent the exchange correlation interactions. Screened hybrid functionals yields an accuracy in par with standard(full-range) hybrids\cite{heyd2005energy,brothers2008accurate,ernzerhof1999assessment,barone2011accurate,janesko2009screened,moussa2012analysis,harb2011origin} but with  reduced computational cost\cite{garza2016predicting,he2017assessing,paier2006screened,marsman2006erratum,henderson2007importance,henderson2008assessment}. When applied to solids, screened hybrids has demonstrated tremendous success in predicting the band gaps of semiconductors and insulators\cite{muscat2001prediction,bredow2000effect} with significantly smaller errors than pure density functional theory (DFT) calculations\cite{heyd2004assessment,he2017assessing,paier2006screened,marsman2006erratum,henderson2007importance,henderson2008assessment}. 

While designing a hybrid functional, the choice of mixing fraction($\alpha$) is crucial to the bonding character, ordering and alignment of bands, and thus in the  electronic and dielectric properties of the material. At an operational level, the parameters that define the screened hybrid functional form a two dimensional space spanned by the range separation parameter(screening parameter ($\omega$) and the fraction of Fock exchange ($\alpha$)\cite{moussa2012analysis}. Given the variety of systems, there is no fixed, universal combination of these parameters ($\omega, \alpha $) that leads to predictions with satisfying accuracy. Mixing fraction is often fixed based on the dielectric constant of the materials\cite{gerosa2017accuracy,skone2014self,gerosa2015electronic} following the interpretation of dielectric constant as inverse screening\cite{garza2016predicting}. Based on fitting a large number of molecular species to atomization energies the fraction of Fock exchange $\alpha $ and the screening parameter $\omega $ were originally set to 0.25 (25\%) and $0.2\ \text{{\AA}}^{-1}$, respectively, for standard HSE06 functional\cite{becke1993new,perdew1996rationale}. In the present investigation,  we maintain the screening parameter at $\omega = 0.20$(corresponding to a screening length $r_s = 2/\omega = 10\ \text{{\AA}}$). However, in the spirit of Ref. \cite{janesko2009screened}, Ref. \cite{ko2016performance} and Ref. \cite{zhang2005theoretical} $\alpha$ is tuned and optimized such that modified HSE06 functional yield a DFT bandgap close to experimental gap. The DFT wavefunction of modified HSE06 functional so obtained with the optimized $\alpha$ is chosen as initial state for MBPT($G_0W_0$+BSE) calculations.

In the single-shot $G_0W_0$ approximation, the orbitals $\ket{nk}$ of well converged ground state DFT  with HSE06 exchange-correlation functional is used to compute QP energies $E_{QP}^{nk}$ as first order corrections. Optical properties were calculated by solving BSE for two particle-Green's functions whose kernel includes local field effects as well as screened electron-hole interaction\cite{hanke1979many,onida2002electronic,sander2015beyond}. In most cases it is sufficient to solve BSE in Tamm-Dancoff approximation\cite{leng2016gw,dancoff1950non,tamm1991relativistic} and corresponding VASP implementation has been made use of in our work. In the BSE routine as implemented in VASP, the screened exchange calculated in preceding $G_0W_0$ calculation are used to approximate the electron-hole ladder diagrams\cite{shishkin2006implementation,shishkin2007accurate,shishkin2007self,fuchs2007quasiparticle}. From the solution of the BSE the dielectric function of anatase \ce{TiO2} and oscillator strengths of optical transitions for incident polarization parallel and perpendicular to the crystallographic $c$ axis is computed. Further, the excitation mechanism is explained by examining the exciton amplitude projections on the quasiparticle bandstructure\cite{wang2017electronic,bokdam2016role}. Based on the analysis, properties such as the nature of photo-generated charge carriers, their origin, spatial localization in the anatase \ce{TiO2} are also presented.  
  
\section{\label{sec:results} Results and Discussion}
\subsection{\label{sec:optimizing_setup}Optimization of HES06-DFT electronic structure}

\begin{figure}
	\centering{}
	\includegraphics[width = 0.65\textwidth]{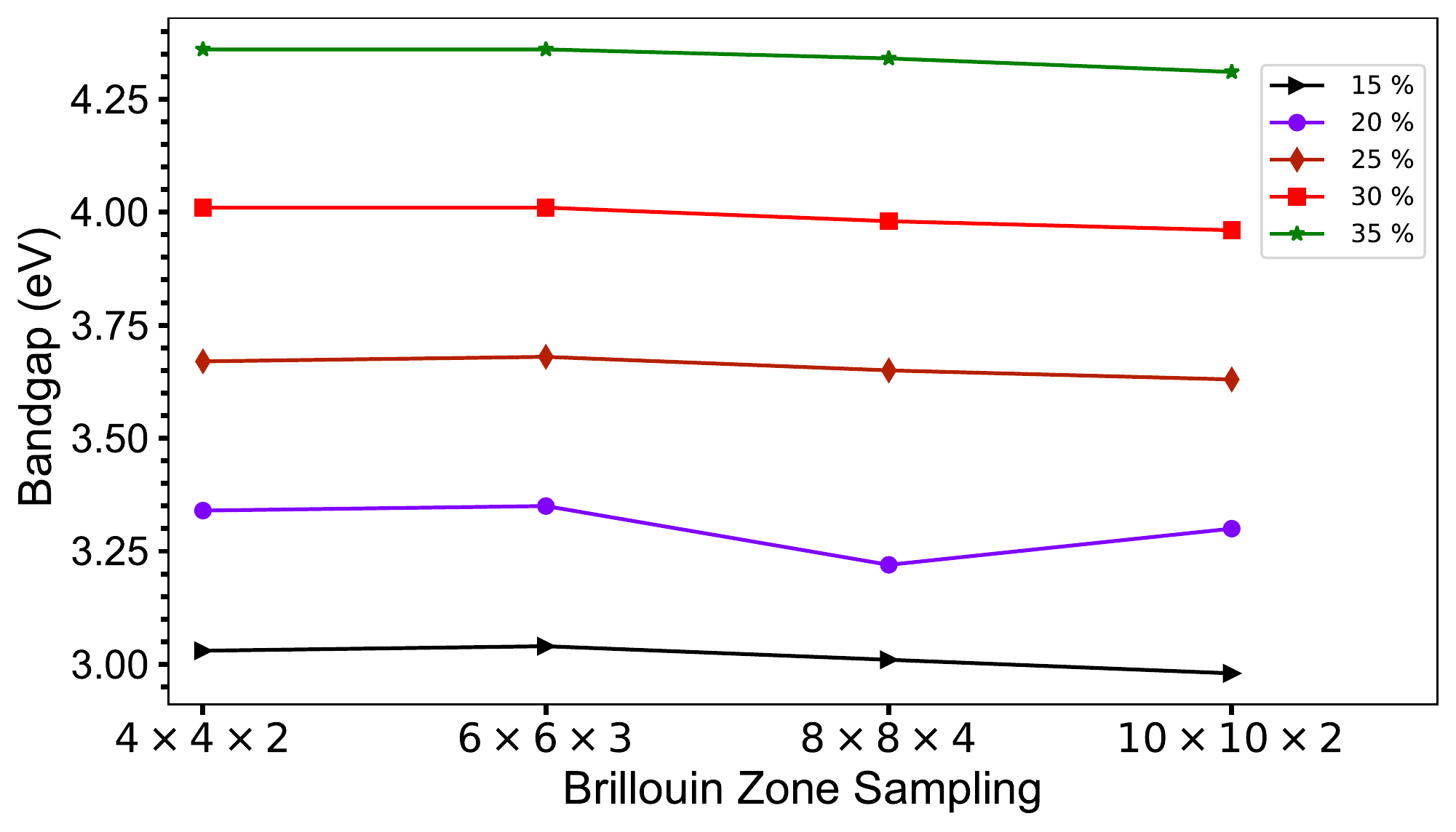}
	\caption{The convergence of HSE06 bandgap for various values of exchange fraction ($\alpha$) with respect to Brillouin zone sampling size. Each curve correspond to $\alpha$ values from 15\% to 35\% with an interval of 5\%. The HSE06($\alpha$) gap is well converged to within 0.05 eV with respect to Brillouin zone sampling. For $6\times 6\times 3$ kpoint grid, as $\alpha$ increases band gap also increases at an approximate rate of 3 eV per 5\% increase in the $\alpha$, which  is represented quantitatively by a straight line $E_g(eV) = 0.066 \alpha(\%) + 2.038 $. The intercept of the fitted straight line closely matches the PBE band gap(2.11 eV). Due to the relative closeness of HSE06(20) gap to the experimental result, HSE06(20) wavefunction is chosen as zero order starting point for many-body perturbation theory calculations} 
	\label{fig:HSE_Bandgap_vs_KPT}
  \end{figure}

The search for optimal DFT wavefunction with HSE06 functional is carried out by varying the fraction of exact exchange ($\alpha$) from 15\% to 35\% with an interval of 5\%. This range of $\alpha$ for hybrid functionals has been shown to yield very good results for large class of systems\cite{marques2011density,zhang2005theoretical,ccelik2012range,de2002effect}. HSE06 ground state corresponding to each $\alpha$ is henceforth referred to as HSE06($\alpha$). For each $\alpha$ the convergence of HSE06 bandgap is ensured by testing the convergence on 4 different k-point sets. The details of convergence study is presented in the figure \ref{fig:HSE_Bandgap_vs_KPT}. As is evident from the figure \ref{fig:HSE_Bandgap_vs_KPT}, in spite of the change in the density of kpoint sampling HSE06 gaps are seen to be converged to within 0.05 eV. Hence kpoint set for all further calculations has been fixed to a $6 \times 6 \times 3 $ grid. Typically, this density of kpoint sampling is more than sufficient to accurately represent the screened exchange interaction in metals and semiconductors, where as, dealing with bare exchange requires at least a $(12 \times 12 \times 12)$ grid\cite{paier2006screened}. This demonstrates the computational efficiency one can achieve when screened hybrid exchange functionals are being used. For $6 \times 6 \times 3 $ the band gap increases from 3.04 eV to 4.36 eV as the fraction of exchange is increased from 15\% through 35\%. This apparent widening of HSE06 gaps in linear proportion to the increase in $\alpha $ ($\approx$ 3 eV for every 5\% increase in $\alpha $) is consistent with the trend observed for hybrid functionals elsewhere\cite{ko2016performance, vines2017systematic,marques2011density,zhang2005theoretical}.   

The best converged result for the HSE06($\alpha$) band gap, compared to  experimental reports (3.2 eV), is obtained for 20\% exact exchange fraction with $6 \times 6 \times 3 $ grid sampling the Brillouin zone, which is $3.34 \pm 0.05$ eV. This setting is maintained for the rest of the calculations, more importantly for optics calculations, as it reproduces the correct bandgap and band dispersion in anatase \ce{TiO2}. For the rest of the manuscript we would use the notation HSE06(20) to represent the modified HSE06 functional with $\alpha = 20\%$ in order to distinguish it from the standard HSE06 functional.

\begin{figure}
	\centering{}
	\includegraphics[width=0.9\textwidth, keepaspectratio]{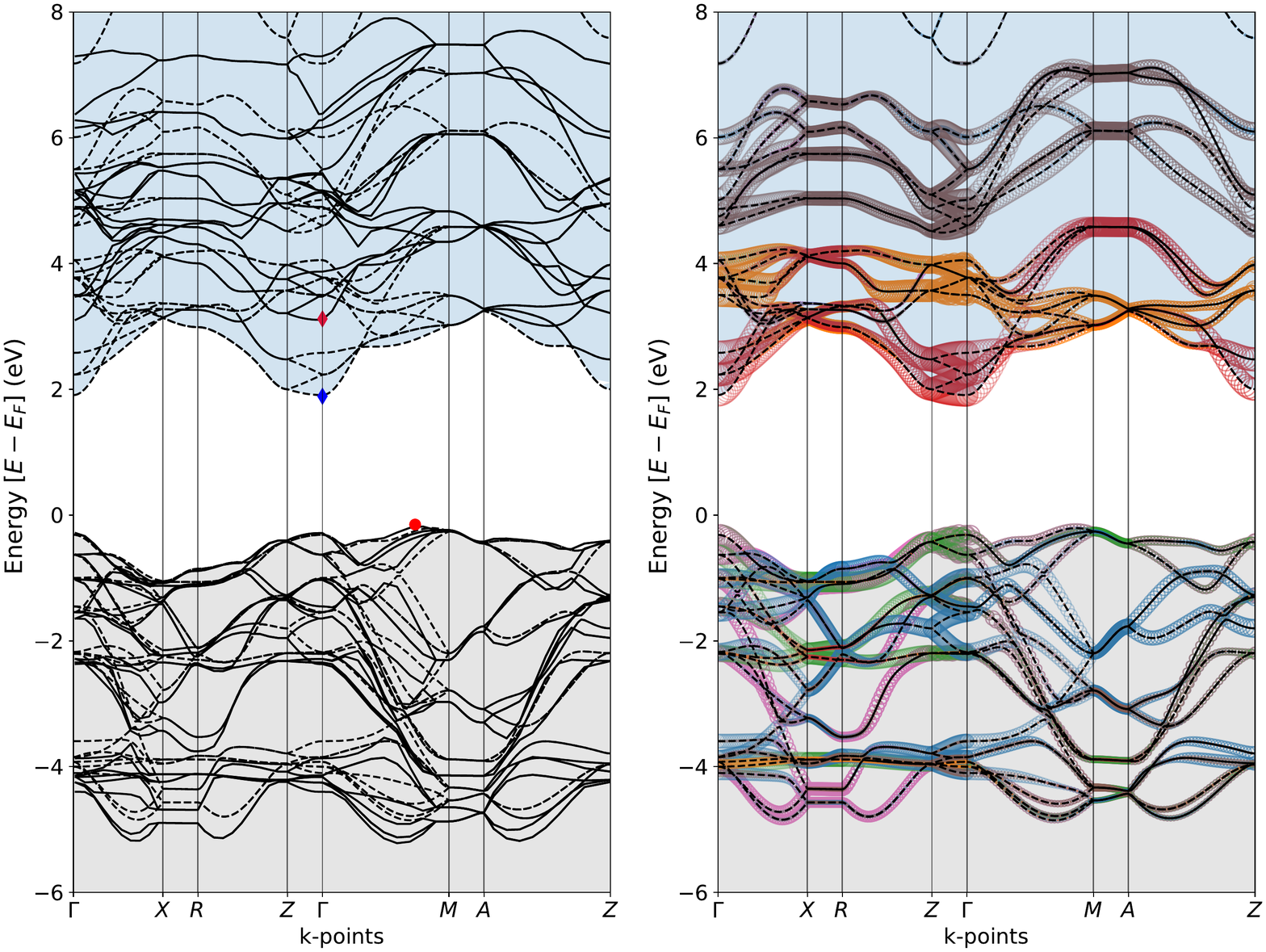}
	\caption{The HSE06(20) and GGA(PBE) band structure calculated along the $\Gamma-X-R-Z-\Gamma-M-A-Z $ high symmetric directions in the irreducible Brillouin zone (IBZ) of anatase \ce{TiO2}. Dotted and solid black lines represent the PBE and HSE06(20) bands, respectively. The energy reference is taken to be the Fermi energy. In the left panel, the red and blue diamonds label the conduction band minimum as obtained in HSE06(20) and GGA(PBE) bandstructure, respectively. The red circle labels the valence band maximum, which is the same with both GGA and HSE06(20). The value of bandgap as obtained with HSE06(20) and GGA(PBE) are both indirect, which are 3.34 eV and 2.11 eV, respectively. Bands are generally dispersive indicating at all regions of the Brillouin zone showing the covalent nature of bonding between \ce{Ti} and \ce{O} ions. In the right panel, O-$2p$ and Ti-3$d$ orbitals are projected onto the PBE bandstructure in a fatband form. In the figure circles projected on the bands represent O-$p_x$ (pink), $p_y$ (green), $p_z$ (blue) and Ti-$d_{xy}$ (red), $d_{yz}$ (orange), $d_{xz}$ (gray), $d-t_{2g}$ (brown) orbital contributions to the bandstructure. Topmost valence band and lowest conduction band, which are nearly parallel along the $\Gamma - Z$ direction, provide joint density for intense optical transitions between O-$p_x$ and $p_y$ states in the valence band to Ti-$d_{xy}$ in the conduction band.}
	\label{fig:DFT_HSE_Bands}
  \end{figure}

Fig. \ref{fig:DFT_HSE_Bands} presents the DFT bandstructure and the density of states of anatase \ce{TiO2} computed using the HSE06(20) functionals along the high symmetric directions in the irreducible Brillouin zone of \ce{TiO2}. For comparison, bandstructure obtained with local/semi-local GGA(PBE) functionals is also superposed in the figure \ref{fig:DFT_HSE_Bands}. It is interesting to note that the dispersion of conduction and valence bands near the respective band edges remain similar under both GGA and HSE06(20) functional description\cite{S_fig1}. The role of HSE06(20) functional on eigenvalues appear nearly like a rigid upward shift of conduction bands, almost uniformly across all the bands and kpoints, from PBE counterparts. With the application of HSE06(20) functionals, the conduction band opens up by 1.23 eV, while the valence band remains unaffected. With conduction bands shifting up, the overall fundamental gap of pristine anatase \ce{TiO2} opens up to 3.34 eV. The computed bandgap is indirect with CBM at $\Gamma $ point and VBM at 0.91$\Gamma \rightarrow M$\cite{mo1995electronic,zhang2005theoretical}, consistent with the literature \cite{boonchun2016energetics,landmann2012electronic,chiodo2010self}. We observe that, the HSE06(20) DFT, in general, predicts bandgap which is more consistent with experimental data than LDA/GGA(PBE) functionals.

Compared to HSE06(20), standard HSE06(with $\alpha$ = 0.25) yields the fundamental gap at 3.68 eV in our calculation, which is overestimated. Similar overestimation tendencies have also been shown by other studies on \ce{TiO2} utilizing standard hybrid functionals; which are 3.60 \cite{landmann2012electronic,deak2012quantitative}, 3.57\cite{dou2013comparative}, 3.58\cite{deak2011polaronic}, 3.59\cite{gerosa2015electronic}, 3.60\cite{arroyo2011dft+}, and 3.89 eV\cite{mattioli2010deep}. Another study bench-marking the performance of various functionals on large class of transition metal oxides and dichalcogenides reports a gap of 3.38 eV with standard HSE06 functional\cite{borlido2019large}. As a measure to remedy the overestimation of bandgap of \ce{TiO2} by standard hybrid functionals, modified $\alpha$ has as well been tested in extant studies. 20\% exchange fraction in the HSE06 functional has been tested earlier for anatase \ce{TiO2} by \citeauthor[]{janotti2010hybrid}. They obtain a bandgap of 3.05 eV, which is slightly underestimated\cite{janotti2010hybrid}. In another study, \citeauthor[]{ccelik2012range} found an indirect bandgap of 3.20 eV in the $X-\Gamma $ direction by employing HSE functional with a mixing fraction of 0.22\cite[]{ccelik2012range}.

  \begin{figure}
	\centering
	\includegraphics[width=0.7\textwidth, keepaspectratio]{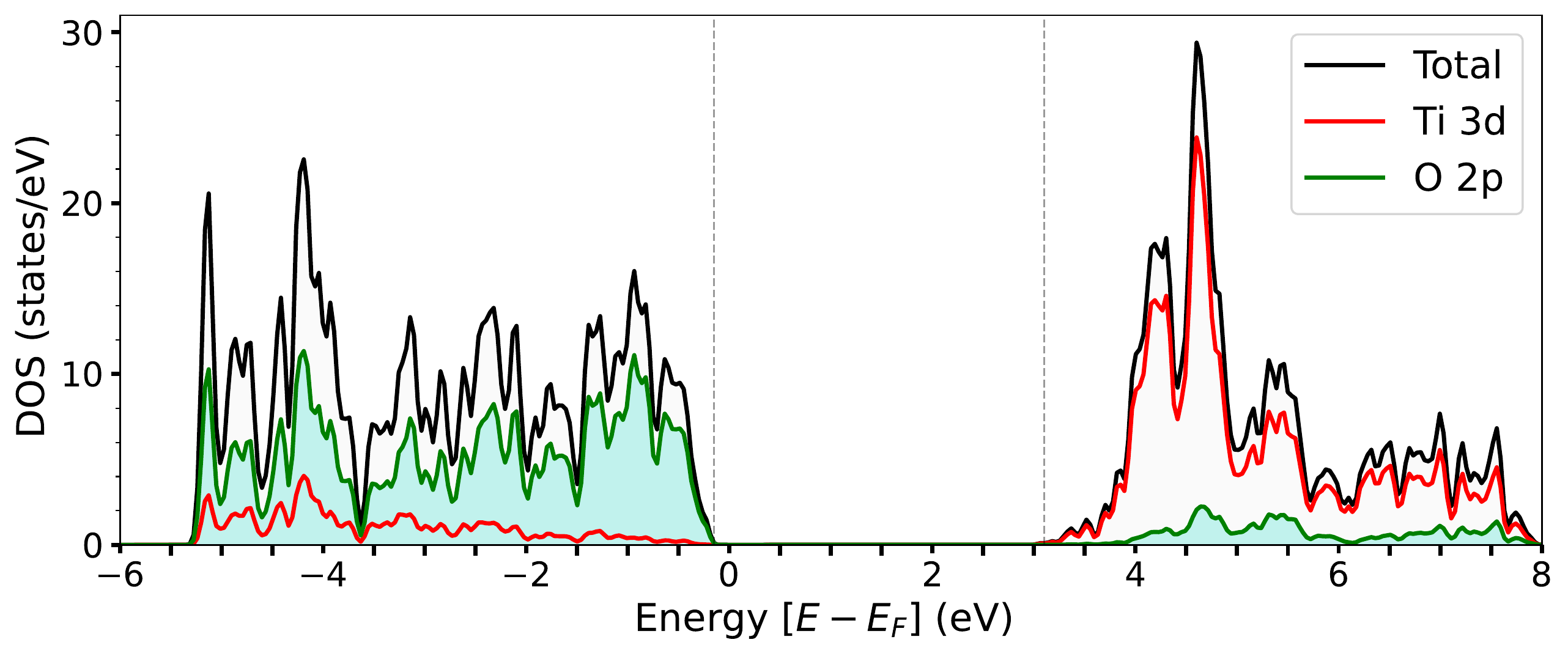}
	\caption{The total and atom projected-angular momentum decomposed density of states showing the Ti-3\textit{d} and O-2\textit{p} contributions to the local density of states: Energies in DOS are reported with reference by setting Fermi energy as zero. In the DOS, valence bands and conduction bands predominantly composed of O-2\textit{p} and Ti-3\textit{d} states respectively. However, there are hybridization of Ti-3\textit{d} and O-2\textit{p} orbitals in both the regions. Ti 3\textit{d} states in the conduction band region is split into two regions, lying above and below 4.5 eV, as a result of degeneracy of Ti-3\textit{d} orbitals being lifted due to the distorted \ce{TiO_6} octahedra. States below 4.5 eV constitute the $t_{2g}$ states and those above form $e_g$ based on the crystal field splitting induced by the distorted octahedral coordination of Oxygen ions around Ti.}
	\label{fig:HSE_DOS}
  \end{figure}
  
  HSE06(20) density of states of anatase \ce{TiO2} shown in figure \ref{fig:HSE_DOS} and the fatband representation of O-$2p$ and Ti-3$d$ orbitals projected on to the PBE bandstructure (figure \ref{fig:DFT_HSE_Bands}, right panel) reveals that the top region of the valence band is occupied by O-$p$ orbitals. Conduction bands populated primarily by the Ti-3\textit{d} like orbitals, shows a splitting mainly into two groups, one lying above and one below 4.5 eV. This splitting is a result of lifting of the degeneracy of Ti-3\textit{d} orbitals due to the distortion of \ce{TiO_6} octahedra, typical to anatase\cite{lawler2008optical,asahi2000electronic}. The coordination of Oxygen ions around Titanium ions form a distorted octahedral crystal field inducing splitting of Ti-3\textit{d} orbitals into triply degenerate $t_{2g}$ (states below 4.5 eV) and doubly degenerate $e_g$ (states above 4.5 eV)\cite{landmann2012electronic,asahi2000electronic,lucovsky2001electronic,lawler2008optical}. The density of states shown in Ref. \cite{S_fig2} reveals that the joint contribution of Ti-3\textit{d} and O-2\textit{p} orbitals near the band edges, coupled with strong dispersion of bands  is an indication of hybridization of 0-2\textit{p} and Ti-3\textit{d} orbitals. Hybridization of orbitals in the vicinity of Fermi level leads to the formation of covalent Ti-O bonds\cite{asahi2000electronic,fischer1972x,modrow2003calculation}. Covalent nature of the bonds has been established also by determining the charge enclosed within the Bader charge volume\cite{tang2009grid} of respective ions. From the Bader charge analysis we arrive at Bader-charge based ionic charge states of \ce{Ti^{2.26+}} and \ce{O^{1.12-}}. The departure of ionic charge states from the expected nominal ionic valences of +4 and -2 for \ce{Ti} and \ce{O}, respectively, is an indication of strong covalent character of the \ce{Ti-O} bond. It has been established here that the involvement of $d$ orbitals in the electronic structure of \ce{TiO2} brings about a significant degree of covalent character and distortion of otherwise atom-like charge surrounding the Ti ion. As it is evident from the preceding discussion that, HSE06(20) very accurately reproduces the character of density of states as has been predicted in the literature\cite{asahi2000electronic,lawler2008optical,landmann2012electronic}. We, by this analysis, have demonstrated that HSE06(20) functional captures the effect of Ti-3\textit{d} electrons in the electronic structure, making it a  good starting point for $G_0W_0$ and BSE calculations for \ce{TiO2} anatase.

\subsection{\label{sec:QP_corrections}$G_0W_0$ Quasiparticle corrections}
\begin{figure}[ht]
	\centering{}
	\includegraphics[width=0.7\textwidth, keepaspectratio]{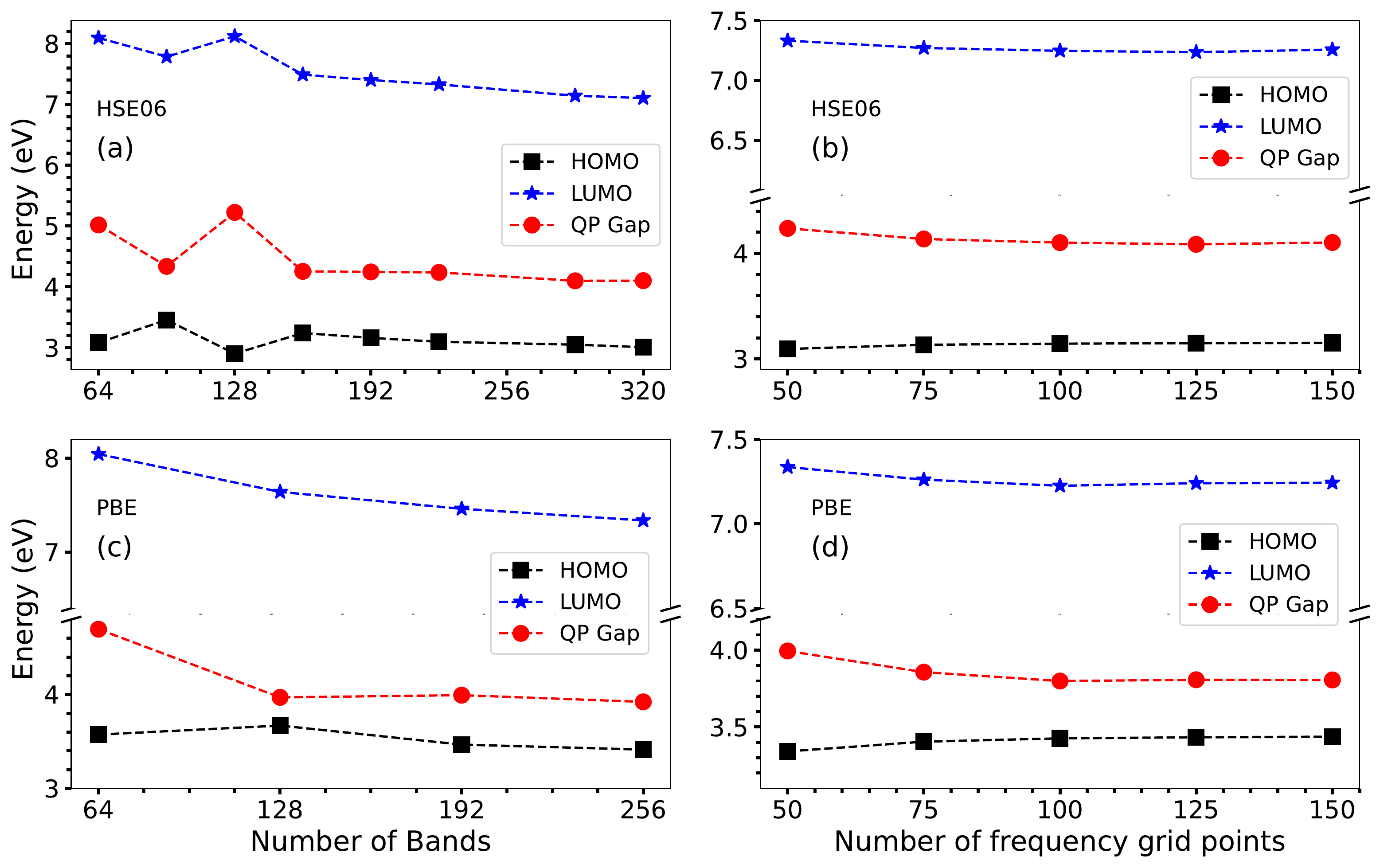}
	\caption{Convergence of $G_0W_0$ HOMO and LUMO energies, and bandgaps with respect to number of bands and number of frequency grid points for PBE and HSE06(20) starting wavefunction. Convergence of QP gap is faster compared to absolute HOMO and LUMO energies. QP gaps reported in this article is well converged to within 0.01 eV. Further, the convergence of absolute energies and QP energies better than 0.01 eV would require even more number of bands and denser k-point sampling. Convergence is moderately similar for both GGA and HSE06 starting wavefunctions}
	\label{fig:GW_Bandgap_Covergence}
  \end{figure}

The quasiparticle(QP) electronic structure is determined within the framework of MBPT using the single-shot $G_0W_0$ approximation. $G_0W_0$ QP corrections are computed on DFT eigenvalues obtained with HSE06(20) functionals(referred to as $G_0W_0$-HSE06(20)). In order to justify the comparison of $G_0W_0$ electronic structure with the experimental outcomes, it is imperative to ensure that the calculation is well converged\cite{karlicky2014band,shishkin2007self,qiu2013optical,qiu2013optical,shishkin2006implementation,kang2010quasiparticle}. We carefully examine the convergence of the $G_0W_0$ calculations with respect to the number of empty states(virtual orbitals) and number of frequency points in the real space for the response functions. For all the convergence studies, the target quantity is the bandgap which is known to converge much faster than the absolute quasiparticle energies due to cancellation effects\cite{kang2010quasiparticle}. The cut-off for basis set for $G_0W_0$ response function is set at rather conservative value of 150 eV for all calculations performed. The details of the convergence study is summarized in figure \ref{fig:GW_Bandgap_Covergence}. Also shown in the figure \ref{fig:GW_Bandgap_Covergence} is a comparison of convergence behavior when PBE and HSE06(20) wavefunctions are used as initial wavefunction for $G_0W_0$ QP corrections. Within the range of values of convergence parameters used in this work, QP gaps as well as  HOMO and LUMO energies show similar rate of convergence irrespective of the starting wavefunction. Quasiparticle band gap is converged to within 0.01 eV when 256 bands (48 occupied and 208 virtual orbital), 100 points on the frequency grid and a $6\times6\times3$ gamma centered kpoint grid are used. For even better convergence, one has to use larger energy cut-off, denser kpoint sampling and more number of empty states which would make the calculation practically not feasible due to forbidding memory requirements\cite{shih2010quasiparticle}.  

Our best converged $G_0W_0$ quasiparticle calculation on top of modified HSE06 (20) wavefunction predicts a quasiparticle (QP) gap of 4.10 eV(indirect)/4.14 eV(direct) for pristine anatase \ce{TiO2}. This value is in reasonable agreement  with prior studies that  report $G_0W_0$ QP gap obtained over hybrid functional starting points. Two of the most important contributions in this direction are by \citeauthor[]{kang2010quasiparticle}\cite{kang2010quasiparticle} and \citeauthor[]{landmann2012electronic}\cite{landmann2012electronic}, and both predicted  band gap of 4.05 eV with HSE06+$G_0W_0$ and HSE+$G_0W_0$, respectively. Rest of the $G_0W_0$ calculations on \ce{TiO2} are performed on top of GGA or GGA+U DFT wavefunction starting points, and the most important results are compiled in Table \ref{table:bandgap_compilation}. As it is evident from the Table \ref{table:bandgap_compilation}, $G_0W_0$ QP gaps computed with GGA and GGA+U DFT wavefunction starting points lie within the range from 3.3 to 3.92 eV and they are all lower than the QP gaps obtained on hybrid starting points\cite{thatribud2019electronic,chiodo2010self,landmann2012electronic,kang2010quasiparticle,patrick2012gw}. The huge variability of $G_0W_0$ QP gaps in the table \ref{table:bandgap_compilation} shows the poor consensus in the literature regarding the calculated QP bandgap of anatase \ce{TiO2}, revealing its strong sensitivity to computational setting and initial structural geometries.

\begin{table*}[ht] 
	\caption{$G_0W_0$ Quasiparticle gap ($E_{gap}^{G_0W_0}$ ) and BSE optical gaps ($E_{gap}^{Opt}$) computed over LDA/GGA, GGA+U and various hybrid functional starting wavefunction reported so far in the literature. All energies are in eV. The symbols I and D in parenthesis of QP gaps denotes if the band gaps are indirect or direct, respectively.}
\begin{ruledtabular}
	\begin{tabular}{*5c}
		% \toprule
		Starting Point & \multicolumn{2}{c}{$G_0W_0$ QP Gap}& \multicolumn{2}{c}{Optical Gap} \\
		\cline{2-3}\cline{4-5} \\[-3.5ex] 
		 & $E_{gap}^{G_0W_0}$ (eV) & Ref.   & $E_{gap}^{Opt}$ (eV)  & Ref. \\ [.5ex]
                \hline \\[-3.5ex]
                HSE06(20)                          & 4.1(I)/4.14(D)       & Present work                       & 3.911  & Present work   \\
		HSE06                                & 4.05(I)              & \cite{kang2015influence}   & 3.45  & \cite{basera2019self}  \\
		HSE06                                & 3.89(I)              & \cite{dou2013comparative}  &   &                                         \\  
		HSE                              & 4.05(I)              & \cite{landmann2012electronic}        &  &   \\ 
	        \cline{1-5} \\[-3.5ex]									   	      
						   & 3.8(I)/3.838(D)                  & Present work                      & 3.745    & Present work  \\
						   & 3.56(I)/4.14(D)      & \cite{kang2010quasiparticle}    & 3.57  & \cite{landmann2012electronic}  \\
						   & 3.83(I)/4.29(D)      & \cite{chiodo2010self}           & 3.90  & \cite{chiodo2010self}  \\
						   & 3.73(I)/3.78(D)      & \cite{landmann2012electronic}   & 3.76  & \cite{baldini2017strongly} \\
		LDA/GGA            & 3.73(I)              & \cite{gerosa2015electronic}     & 4.0   & \cite{lawler2008optical,kang2010quasiparticle} \\               
						   & 3.7(I)               & \cite{patrick2012gw}            & 4.5   & \cite{landmann2012electronic}  \\ 
						   & 3.791(I)             & \cite{thulin2008calculations}   &   &   \\ 
						   & 3.61(I)/3.92(D)      & \cite{baldini2017strongly}      &   &   \\ 
						   & 3.92(I)              & \cite{thatribud2019electronic}  &   &   \\ 
						   & 3.5(I)/3.8(D)        & \cite{giorgi2011excitons}       &   &   \\
                \cline{1-5} \\[-3.5ex]
		GGA+U				   & 3.27                 & \cite{patrick2012gw}            &                  &                                         \\
		% \bottomrule
		\end{tabular}
\end{ruledtabular}
\label{table:bandgap_compilation}
\end{table*}
Considering the tendencies of HSE06 functionals to overestimate the $G_0W_0$ QP gap of anatase \ce{TiO2}, there seems to be no real advantage of performing $G_0W_0$ over hybrids in terms of accuracy of QP gaps. However, it can be shown that $G_0W_0$ calculation on top of HSE06 wavefunction captures quasiparticle nature of electronic excitation and excitonic character of optical excitation which the GGA or DFT+U starting points fail to do\cite{landmann2012electronic}. This is evident from the amount of QP correction we obtain over HSE06($\alpha$) and PBE starting wavefunction. We obtain a $G_0W_0$ QP correction with respect to DFT indirect gap (2.11 eV) of 1.69 eV for anatase \ce{TiO2} over PBE starting point wavefunction while the same over the HSE06(20) indirect bandgap(3.34 eV) amounts to 0.76 eV only. With HSE06(25) and HSE06(30) starting points, QP corrections are 0.49 eV and 0.24 eV, respectively, over corresponding HSE06($\alpha$) bandgap (Table \ref{table:exciton_binding_energy_aexx}). As the exchange fraction($\alpha$) is increased the $G_0W_0$ correction for calculations starting from respective modified HSE06($\alpha$) bandstructure tend to vanish. Increasing $\alpha$ makes the functional better at capturing quasiparticle nature of the excitation, and remedies the self-energy problem of PBE functionals. However, this would lead to further overestimation of fundamental gap of anatase \ce{TiO2}.

\subsection{\label{sec:optical_transitions} Optical transitions}
To identify the microscopic nature of the optical excitation, the dielectric function of pristine anatase \ce{TiO2} is evaluated by solving Bethe-Salpeter Equation(BSE). The $W_0$ of the preceding $G_0W_0$-HSE06(20) calculation and a $\Gamma$-centered, un-shifted $6 \times 6 \times 3$  kpoint grid has been employed in the BSE calculation. BSE with HSE06(20) hybrid exchange as used in this work is henceforth referred to as BSE-$G_0W_0$-HSE06(20) in the rest of the paper. The dielectric function is evaluated at 100 frequency points on the real axis. 20 BSE eigenvalues for the incident light polarized parallel and perpendicular to the $c$-axis are computed in the BSE step. Optical transition energies and exciton binding energies are thoroughly converged, each to within 10 meV, by including sufficient number of occupied and virtual orbitals in the BSE calculation\cite{S_fig3}. For the final results presented here, 8 highest occupied and 16 lowest empty bands of the $G_0W_0$-HSE06(20) bandstructure are considered for solving BSE.  

\begin{figure}
	\centering{}
	\includegraphics[width=0.8\textwidth, keepaspectratio]{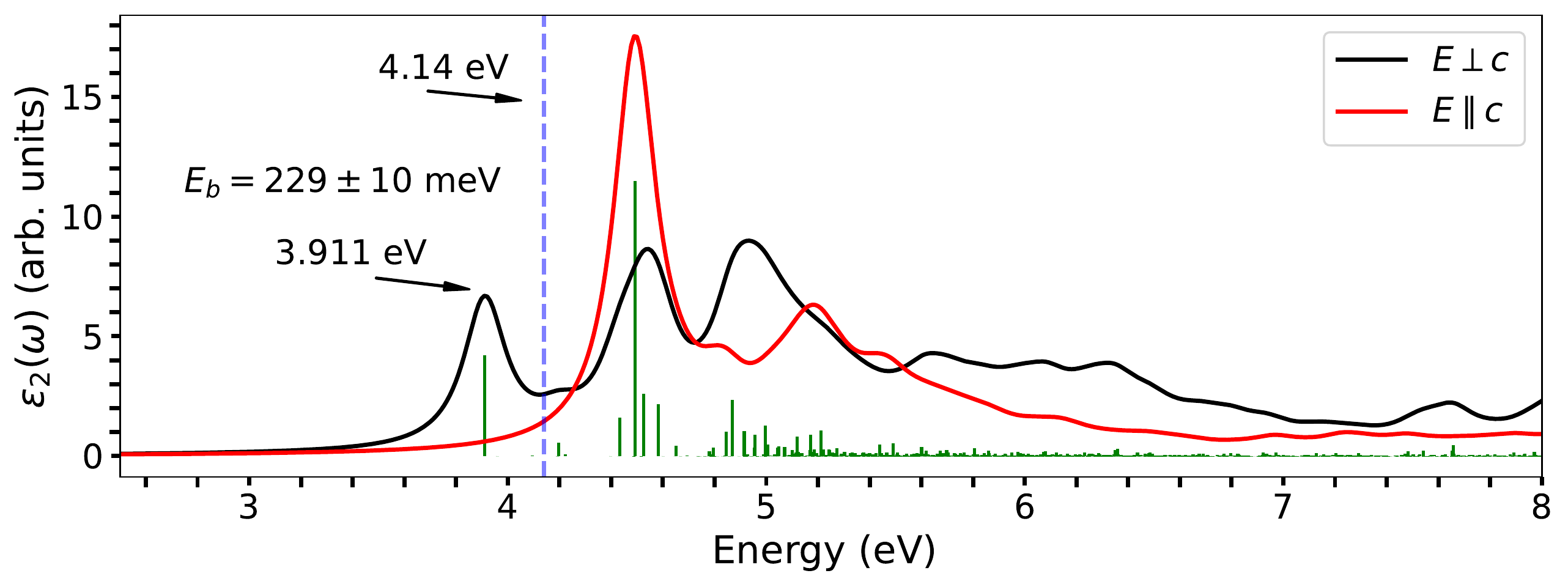}
	\caption{The imaginary part of the dielectric function of anatase \ce{TiO2} computed within the BSE-$G_0W_0$-HSE06(20) set up. In the figure, the optical response is shown for in-plane polarization($E\perp c$)  and direction perpendicular to crystallographic c axis($E\parallel c$) by black and red solid curves, respectively. Green spikes in the figure represent the relative magnitude of oscillator strengths corresponding  to optical transitions for both polarization. First peak in the dielectric function of anatase \ce{TiO2} is an excitonic peak. From the difference between first optical absorption obtained as solution to BSE and corresponding quasiparticle gap obtained in the $G_0W_0$-HSE06(20) calculation(blue, dashed vertical line at 4.10 eV) the exciton binding energy is estimated to be 229 $\pm$ 10 meV .}
	\label{fig:HSE_BSE_Dielectric_HSE_BSE_Oscillator_Strength}
\end{figure}

\begin{figure}
	\centering{}
	\includegraphics[width=0.7\textwidth, keepaspectratio]{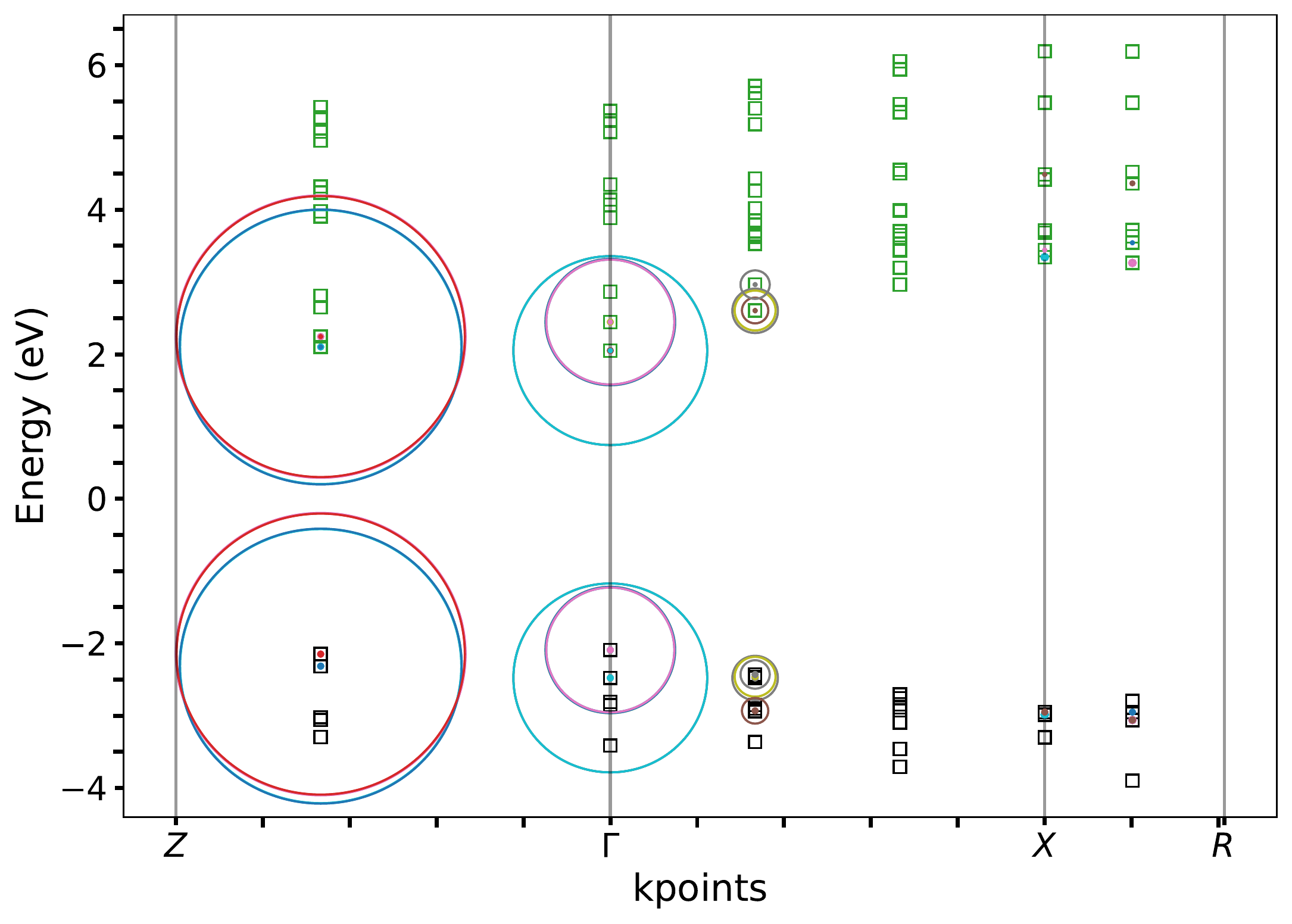}
	\caption{The third BSE eigenvalues of electron-hole pairs (corresponding to first excitonic transition) of anatase \ce{TiO2} is visualized along with the strength coupling between them in fatband style. Black and green open squares in the fatband are BSE eigenvalues in the valence band and conduction band, respectively. Energies in the plot are shifted to make Fermi energy zero. The pairs of circles of same color and radii, one centered at a hole eigenvalue from the valence band and the other at an electron eigenvalue at the conduction band at a given k-vector represent an electron-hole pair. The radius of the circle is an indicative of relative coupling strength of individual exciton. The $\Gamma$ and $Z$ region of the Brillouin zone predominantly contribute to the first excitonic transitions while the rest of the regions contributes only weakly. This clearly shows the localization of excitons in the plane perpendicular to the c-axis of anatase unit cell.}
	\label{fig:Fatband_HSE}
\end{figure}

The computed imaginary part of the dielectric function ($\epsilon_{2}(\omega)$) of anatase \ce{TiO2} along with oscillator strengths of optical transition is presented in figure \ref{fig:HSE_BSE_Dielectric_HSE_BSE_Oscillator_Strength}. Taking the anisotropy associated with the tetragonal symmetry of the anatase lattice into account we resolve the $\epsilon_{2}(\omega)$ into two components-\ $E\perp c$ which is the average over the $x$ and $y$ components, and the $E\parallel c$(the $z$ component). The first peak of $\epsilon_{2}$ occurs for the $E\perp c$ component corresponds to the direct optical gap 3.911 eV. It lies below the related direct quasiparticle bandgap($G_0W_0$-HSE06(20)) of 4.14 eV, confirming that the first direct optical excitation in anatase \ce{TiO2} is dominated by bound excitons. All other excitation are resonant excitation. We compute the binding energy concerning peak at 3.911 eV, i. e., the difference between first direct optical transition and direct quasiparticle gap, to be 229 $\pm$ 10 meV. 

The exciton wavefunction can be expressed as an electron-hole product basis $ \phi' = \displaystyle\sum_{cvk} A^{S}_{vck}\phi_{ck}\phi_{vk}$, where $v$ and $c$ are valence and conduction band states at the $k $ point, and $A^{S}_{vck} $ is the exciton amplitude with the corresponding excitation energy $E^{S}_{exc}$. In order to determine the region of the Brillouin zone important for the optical gap, the eigenstate corresponding to the first exciton ($S = 3$) from the BSE is visualized as fatband in Figure \ref{fig:Fatband_HSE} by plotting circles of radius $\abs{A_{vck}^{3}}$ onto the bandstruture\cite{wang2017electronic,bokdam2016role,kolos2019accurate}. Black and green open squares in the fatband are BSE eigenvalues in the valence band and conduction band, respectively. Centered at the BSE hole(electron) eigenvalue in the valence band(conduction band) at a given $k $ point, pairs of circles of identical colour represent an excitons which contribute most to the first excitonic peak in $\epsilon_{2}$. Only two highest occupied bands ($v = 7, 8$) and two lowest unoccupied bands ($c = 9, 10$) considered for BSE are important for the optical gap. The most important part of the Brillouin zone responsible for the excitonic peak in at 3.911 eV lies along the $\Gamma-Z $ directions. To within the resolution of k-space available in our BSE calculation we can assert that other regions contribute only weakly. Previous BSE calculations have also arrived at a conclusion similar to the present study\cite{chiodo2010self,ataei2017excitonic}. Combining Figure \ref{fig:Fatband_HSE} with the HSE06(20) bandstructure (figure \ref{fig:DFT_HSE_Bands}), these bands are identified to be composed of O-$p_x$, $p_y$ and Ti-$d_{xy}$  orbitals at valence band and the conduction band, respectively. Dispersion of lowest conduction band and the highest valence band in the $\Gamma - Z$ direction are nearly parallel, leading to similar electron and hole group velocities in this region. This provides a large joint density of states for the optical transitions providing stability and strong binding of excitons. Flatness of the bands in the $\Gamma - Z$ direction indicates that orbital interactions in anatase \ce{TiO2} mainly run in the $xy $ plane than in the $z$ direction in the real space\cite{laskowski2009strong}. As excitonic states are also from the same region of the Brillouin zone, this immediately translate into  localization of excitons on the $xy$ plane in real space. Similar observation has been made by Ref. \cite{chen2015anisotropic} and Ref. \cite{baldini2017strongly} wherein it is shown that high degree of localization of bound exciton confine the excitons almost in a single atomic plane(in the $xy$ plane). Besides, the localization of excitons predicted in this work with BSE-$G_0W_0$-HSE06(20) agrees well with the analysis of spatial distribution of the exciton wavefunction in real space by Ref. \cite{chiodo2010self} and Ref. \cite{ataei2017excitonic}. 

\begin{figure}
	\centering{}
	\includegraphics[width=0.8\textwidth, keepaspectratio]{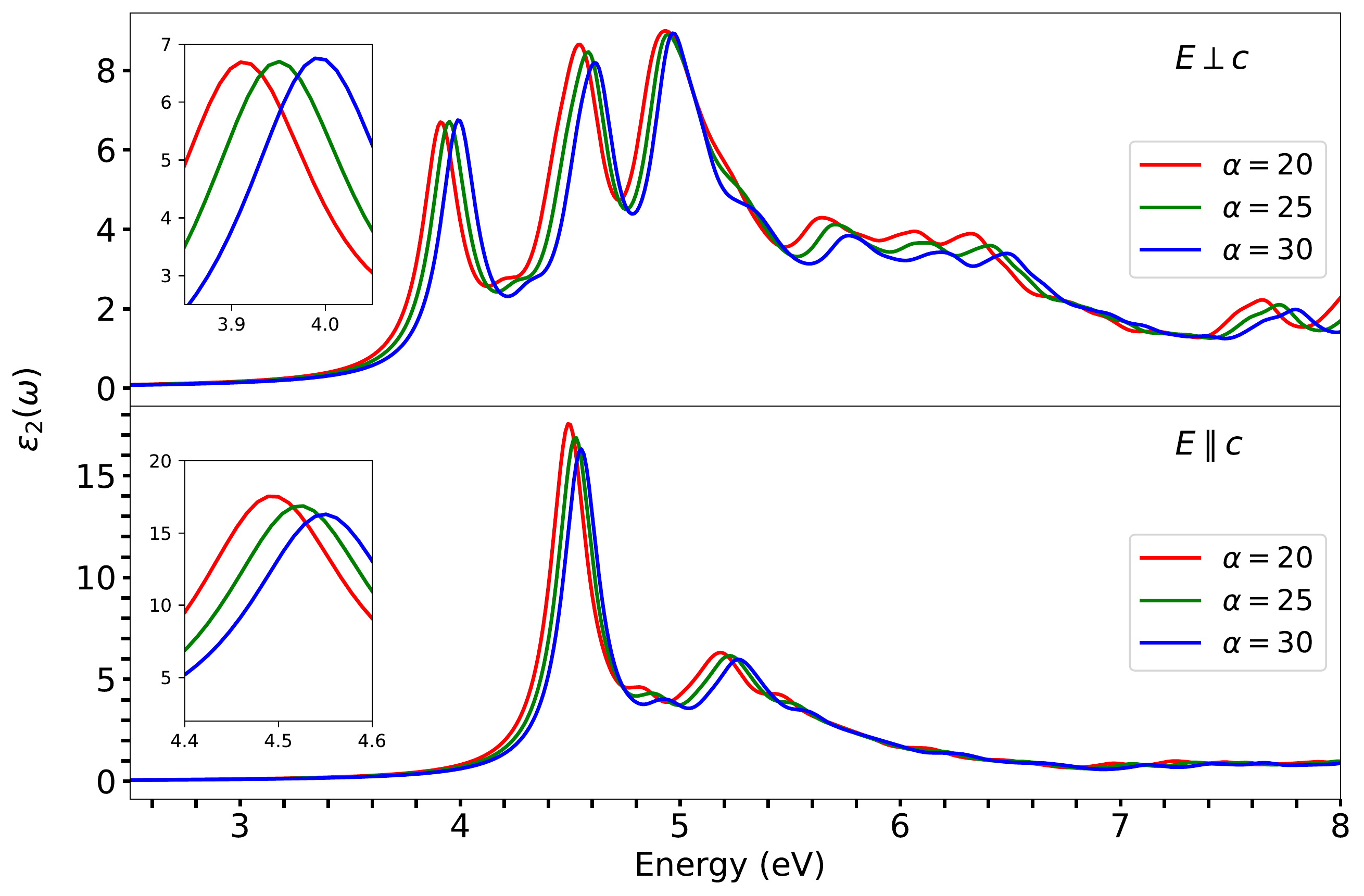}
	\caption{Variation in the imaginary part of dielectric function of anatase \ce{TiO2} for in-plane and perpendicular polarization as a function of the fraction of Fock exchange($\alpha$) in the modified(HSE06) starting point for BSE calculation. The peak correspond to first direct optical transition is blue-shifted as the value of $\alpha$ is increased. In a similar fashion, the excitons become more strongly bound as $\alpha$ is increased.}
	\label{fig:BSE_Spectra_HSE_AEXX_Combined}
  \end{figure}

Figure \ref{fig:BSE_Spectra_HSE_AEXX_Combined} summarizes the influence of fraction of Fock exchange ($\alpha$) in the HSE06($\alpha$) starting wavefunction on the prediction of optical response of anatase \ce{TiO2}. As it is evident from Figure \ref{fig:BSE_Spectra_HSE_AEXX_Combined}, the oscillator strength of optical transitions remains unaffected as the percentage of Fock exchange is varied. Furthermore, the fatband representation of excitonic states in $E\perp c$ direction for HSE06($\alpha$) points to the fact that the Brillouin zone important for the first peak is still along the $\Gamma-Z $ directions, irrespective of $\alpha$\cite{S_fig6}. However, optical gaps and exciton binding energy is strongly affected by it. With HSE06(20) functional starting point we obtain an optical gap of 3.911 eV corresponding to an exciton binding energy of 229$\pm 10 $ meV for this excitation. When $\alpha$ is increased further to 25 and 30 \% optical gap (exciton binding energy) also shifts to higher energies, 3.949 eV (261$\pm 10 $ meV) and 3.992 eV (298$\pm 10 $ meV), respectively. With an increase in the amount of Fock exchange ($\alpha$), the quasiparticle gap ($G_0W_0$ bandgap) becomes wider (Table \ref{table:exciton_binding_energy_aexx}). A larger electronic gap makes the electronic screening less efficient, and consequently the electron-hole interaction is less screened. Therefore, the electron-hole pairs in the interacting levels become more bound. 

In comparison to HSE06($\alpha$) starting point, BSE calculations with PBE wavefunction show that there is no noticeable difference in the oscillator strengths of transitions in this case as well\cite{S_fig5}. This can be very well related to the similarity of DOS and orbital composition for bands lying close to the bandgap in both descriptions. Similarly, the region of the Brillouin zone relevant for first optical peak predicted by PBE and HSE06($\alpha$) starting points are also identical\cite{S_fig7}. On the other hand, due to self-energy and related bandgap problem, the binding energy of excitons for the first peak at 3.745 eV with PBE functional as the starting point is 93$\pm 10 $ meV\cite{S_fig4}. 

  \begin{table*}[ht] 
	\caption{HSE06($\alpha$) indirect gap ($E_{HSEO6}^I(\alpha)$), Indirect and direct quasiparticle gap ($E_{QP}^I, E_{QP}^D$) , quasiparticle correction over indirect bandgap ($E_{QP}^I - E_{HSEO6}^I(\alpha)$), BSE optical gaps ($E_{Optical}^D$) and exciton binding energy of anatase \ce{TiO2} as a function of Fock exchange fraction($\alpha$) in the HSE06($\alpha$) set-up.}
\begin{ruledtabular}
	\begin{tabular}{*8c}
    $\alpha$ & $E_{HSEO6}^I(\alpha)$ & $E_{QP}^I$ & $E_{QP}^I - E_{HSEO6}^I(\alpha) $ & $E_{QP}^D$ & $E_{Optical}^D$  & $EB$ \\[-1.5ex]
       (\%)      & (eV)        & (eV)   & (eV)          & (eV)  & (eV) & (meV)\\ 
    \hline
    20     & 3.34 & 4.10    & 0.76   & 4.14  & 3.911                & 229                         \\ 
    25     & 3.68 & 4.17    & 0.49   & 4.21  & 3.949                & 261                        \\
    30     & 4.01 & 4.25    & 0.24   & 4.29  & 3.992                & 298                        
		\end{tabular}
\end{ruledtabular}
\label{table:exciton_binding_energy_aexx}
\end{table*}

At this point it is important to emphasize that the present BSE calculations are performed for a frozen lattice with no consideration given for electron-phonon interactions. The excitons described in our BSE calculation are screened by the electronic component alone. Ionic relaxation and lattice rearrangements induced by exciton formation\cite{gallart2018temperature} and the influence of intrinsic defects\cite{pascual1978fine,mattioli2010deep} are not taken into account in our calculation. Hence the optical gap computed in our BSE calculations can only be compared with measured direct optical gaps (computational or experimental). In the case of pristine anatase \ce{TiO2}, as evidenced by the large difference between static and optical dielectric constants (static (optical) dielectric constants : 45.1(5.82) for $E\perp c$ and 22.7(5.41) for $E\parallel c$), the lattice relaxations has strong influence on the dynamics of excited charges\cite{gonzalez1997infrared}. Moreover, at finite temperatures, electron-phonon interactions and lattice thermal expansion affect temperature band gap renormalization in anatase \ce{TiO2}\cite{wu2020theoretical}. As a result, 3.2 eV, which is widely considered as the absorption edge of anatase \ce{TiO2} in several experimental investigations\cite{tang1995urbach,hosaka1996uv,hosaka1997excitonic,wang2002optical}, is indeed due to indirect processes\cite{tang1993photoluminescence,tang1994electrical,tang1995urbach,persson2005strong,morgan2009polaronic,deskins2007electron,watanabe2005time,lawler2008optical,baldini2017strongly}. The allowed direct transitions have been previously found by spectroscopic ellipsomerty at 3.8 eV, and 3.79 eV, respectively, by Ref. \cite{wang2002optical} and Ref.\cite{baldini2017strongly} and at 3.87 eV by photoluminiscence as in Ref. \cite{liu2007photoluminescence}. Our estimate of BSE direct optical gap of 3.911 eV computed over HSE06(20) starting wavefunction is in close agreement these experimental data for direct excitation.

Theoretical estimate of optical gaps by BSE calculations of pristine anatase \ce{TiO2} lies in a wide range from 3.45 to 4.5 eV (Table \ref{table:bandgap_compilation}). Extant literature shows that the use hybrid functional as the starting wavefunction for BSE calculation of anatase \ce{TiO2} is not very well explored. One such BSE study reported an optical gap of 3.45 eV\cite{basera2019self} for $E\perp c$ polarization using the standard HSE06 functional as starting wavefunction. This is the  the lowest of all available BSE calculations of pristine anatase \ce{TiO2}, and is underestimated compared to results obtained in present investigation. With PBE functional starting wavefunction, direct optical gap obtained from BSE calculation for anatase single crystals by Ref. \cite{baldini2017strongly}(3.76 eV) yields excellent agreement with  spectroscopic ellipsometric study (3.79 eV), making it the most reliable estimate of optical gap so far. In the present work, optical gap obtained with HSE06(20) starting point (3.911 eV) and that for PBE starting point (3.745 eV) are very similar to each other. Both of these values are closer to experimental value presented in  Ref. \cite{baldini2017strongly} as well. However, because of the smaller  bandgap, the exciton binding energy estimated with PBE starting point (93 meV) is much smaller than its HSE06(20) counterpart (229 meV). In comparison to the value of exciton binding energy derived from experiments (180 meV) and also obtained consistently from frozen-atom BSE calculation (160 meV) in Ref. \cite{baldini2017strongly}, prediction from HSE06(20) starting wavefunction is much closer. Experiments and frozen-atom BSE calculations reported in Ref. \cite{baldini2017strongly} are both performed at 20$^{\circ}$ C. According to Ref. \cite{baldini2017strongly}, at 20$^{\circ}$ C, the combined effect of the temperature dependent lattice expansion and electron–phonon coupling is in fact negligible in pristine anatase \ce{TiO2}. 

\section{\label{sec:conclusion}Conclusion and Outlook}
To conclude, we systematically investigated the quasiparticle electronic structure and and optical excitation of anatase \ce{TiO2} within the framework of many-body perturbation theory (MBPT) by combining the $G_0W_0$ method and the Bethe-Salpeter Equation (BSE). Non-interacting DFT Hamiltonian with a modified version of the HSE06 screened hybrid functional is employed as the staring point for $G_0W_0$ + BSE calculations. Fraction of exact Fock exchange in the HSE06 functional is tuned and the one which yields reasonably accurate ground state electronic structure is chosen as starting point for $G_0W_0$ + BSE simulations. Ground state DFT calculation with HSE06(20) functional yields 3.34 eV as the electronic gap of anatase \ce{TiO2}. It reliably predicts the covalent character of Ti-O bonds, electronic bandstructure, fundamental gap and density of states of \ce{TiO2}. Further, carefully converged $G_0W_0$ + BSE calculation on HSE06(20) starting wavefunction predicts first direct optical excitation of anatase \ce{TiO2} at 3.911 eV, in agreement with experiments that measures direct optical excitation. This optical excitation creates strongly bound excitons with binding energy of 229 $\pm $ 10 meV, with respect to $G_0W_0$-HSE06(20) quasiparticle gap. The projections of excitonic states onto the quasiparticle band structure in a fatband representation shows that the optical transition at 3.911 eV consists of excitons are originating from the mixing of single direct transitions within band pairs running parallel to the $\Gamma -Z $ direction in the tetragonal Brillouin zone. 

The present work highlights the importance of a suitable non-interacting Hamiltonian for the use in quasiparticle MBPT and subsequent BSE calculations. We observe that, irrespective of PBE or HSE06($\alpha = 20, 25, 30 \%) $, the BSE excitonic spectra computed with respective starting points predicts the region of Brillouin zone important for optical gap to be along the $\Gamma -Z $. Nevertheless, as we obtained in the present work, $G_0W_0$ + BSE on PBE functional starting points underestimates the exciton binding energy(93 $\pm $ 10 meV) of anatase \ce{TiO2}. The exciton binding energy as predicted in the present work with HSE06(20) starting wavefunction is more consistent with literature than those with LDA/GGA or DFT+U starting points. This leaves HSE06(20) as an optimal choice for non-interacting Hamiltonian for MBPT calculation in the case of anatase \ce{TiO2}.

Ground state (DFT) and excited state (MBPT) properties computed based on HSE06(20) functional are very encouraging for its continued use and extension into other systems too. This generalization is particularly true in the case of other metal and transition metal oxides where localized $d$ and $f$ electrons strongly influence the electronic structure and optical response. Our study provides a reference to MBPT calculation of doped \ce{TiO2} and with intrinsic defects so that reasonable comparison of physical observable at same level theory can be performed. This MBPT calculation of pristine anatase \ce{TiO2} with HSE06(20) functional staring point will serve as a benchmark for extending calculation into systems  where we cannot exclude the significant role of exchange-correlations in starting wavefunction and $d$ electron localization problem. 

The prediction of strongly bound excitons in anatase \ce{TiO2} and its localization  enabled by BSE calculation on HSE06(20) can potentially help us throw light on some of the challenging and open questions in the literature related to \ce{TiO2}. Characterizing band alignment in anatase and rutile structures with HSE06(20) functionals would deepen our insights on longstanding questions such as why anatase-rutile mixed phases\cite{scanlon2013band,wei2020quantum} are more photo-catalytically active than individual phases. Another challenge is related to the reactivity of different facets in anatase \ce{TiO2}. The particular spatial localization of excitons in the \{001\} plane can partly offer possible explanations to the question as to why the thermodynamically less stable \{001\} facets in \ce{TiO2} is more photo-reactive compared to stable \{101\} ones\cite{yang2008anatase,hu2014synthesis,vittadini1998structure,pan2011true}. The advancement in controlled synthesis of anatase \ce{TiO2} having defined crystal facets at desired crystallographic orientations\cite{selloni2008anatase, yang2008anatase,gong2005reactivity}, harvesting the exciton delocalization in anatase for devices, and nonlinear optical applications has become close to reality. The demonstration of selective and controlled modulation of oscillator strength of excitons by acoustic phonons\cite{baldini2018phonon} and huge exciton shifts by coupling exciton to the optically induced strain pulses\cite{baldini2019exciton} in anatase \ce{TiO2} nanoparticles hold huge promises in this direction. For modelling such physical processes and design further experiments and devices, HSE06(20) hybrid functional based MBPT computations will definitely act as reliable benchmarks and test strategy.

% \clearpage
\bibliography{references.bib}
\newpage
\section*{Supplemental Material}
\begin{figure}[ht]
  \centering{}
  \includegraphics[width=0.8\textwidth, keepaspectratio]{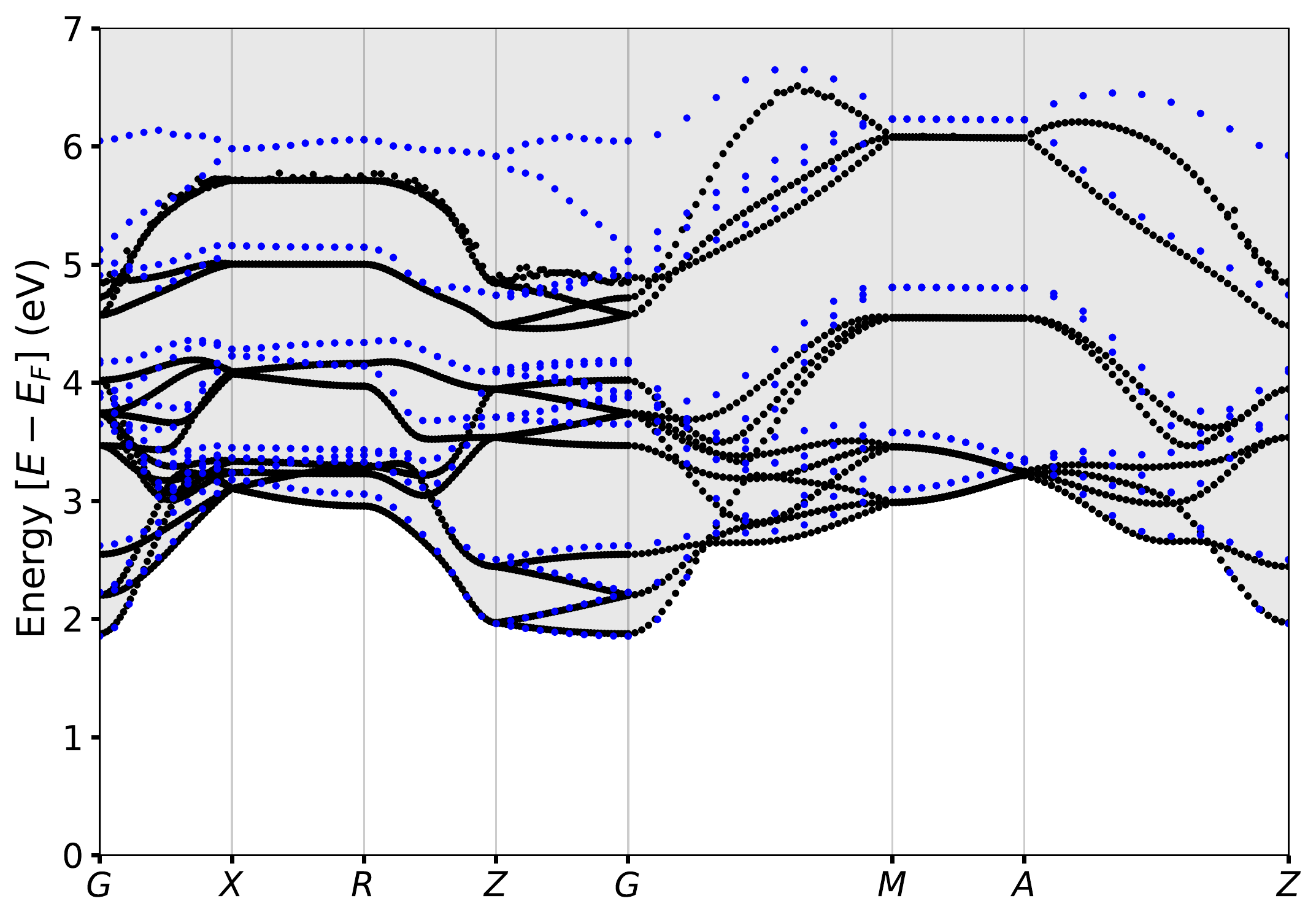}
  \caption{Comparison of conduction band structure under PBE and modified HSE06(20) description. The black and blue dots represents the conduction band energies when PBE and HSE06(20)functionals, respectively, are used . The bands are superposed for comparing dispersion by shifting the HSE06(20) conduction bands uniformly downward so as to match its edge with the PBE conduction edge. Valance band has similar dispersion in both PBE and HSE06(20) description for the exchange correlation, hence not shown here. The general trend we observe is that the dispersion of bands in PBE and HSE06(20) cases agrees with each other in energy ranges very close to the band edges. The agreement worsens further away from the edge. }
  \label{fig:DFT_and_HSE_conduction_bands}
\end{figure}
\begin{figure}[ht]
  \centering{}
  \includegraphics[width=0.7\textwidth, keepaspectratio]{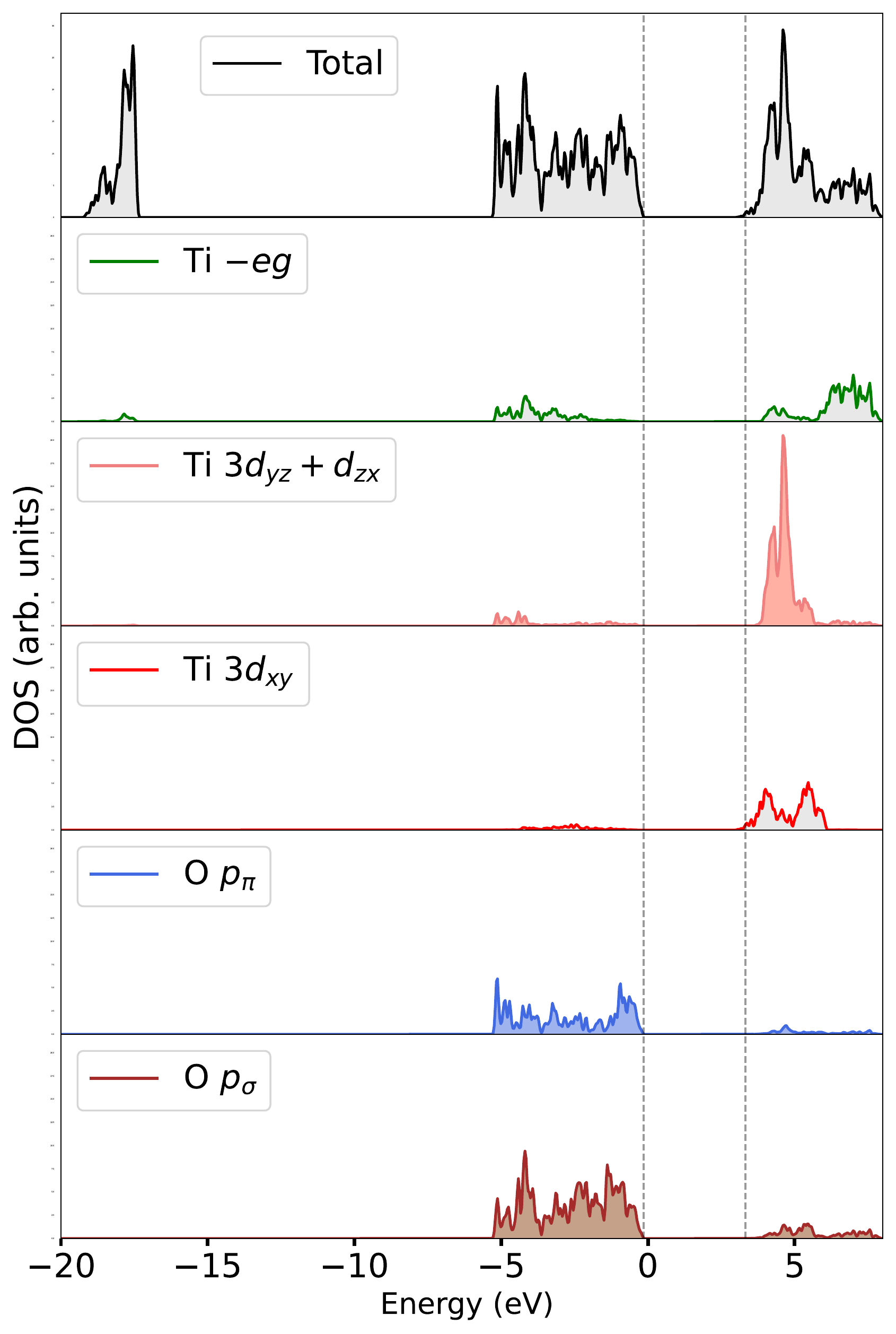}
  \caption{Total and projected density of states of the anatase \ce{TiO_2} structure calculated within the HSE06(20) set up. The Fermi energy is taken as the zero of energy in the bandstructure. The vertical dashed lines on the left and right indicate the valence band maximum and conduction band minimum, respectively. The DOS is decomposed into Ti-$e_g$, Ti-$t_{2g}(d_{yz},d_{zx}$ and $d_{xy})$, O-$p_{\sigma}$ in the \ce{Ti_3-O} cluster plane and O-$p_{\pi}$ out of the \ce{Ti_3-O} cluster plane components. }
  \label{fig:Total_DOS_HSE_vs_GGA}
\end{figure}

\begin{figure}
  \centering{}
  \includegraphics[width=0.8\textwidth, keepaspectratio]{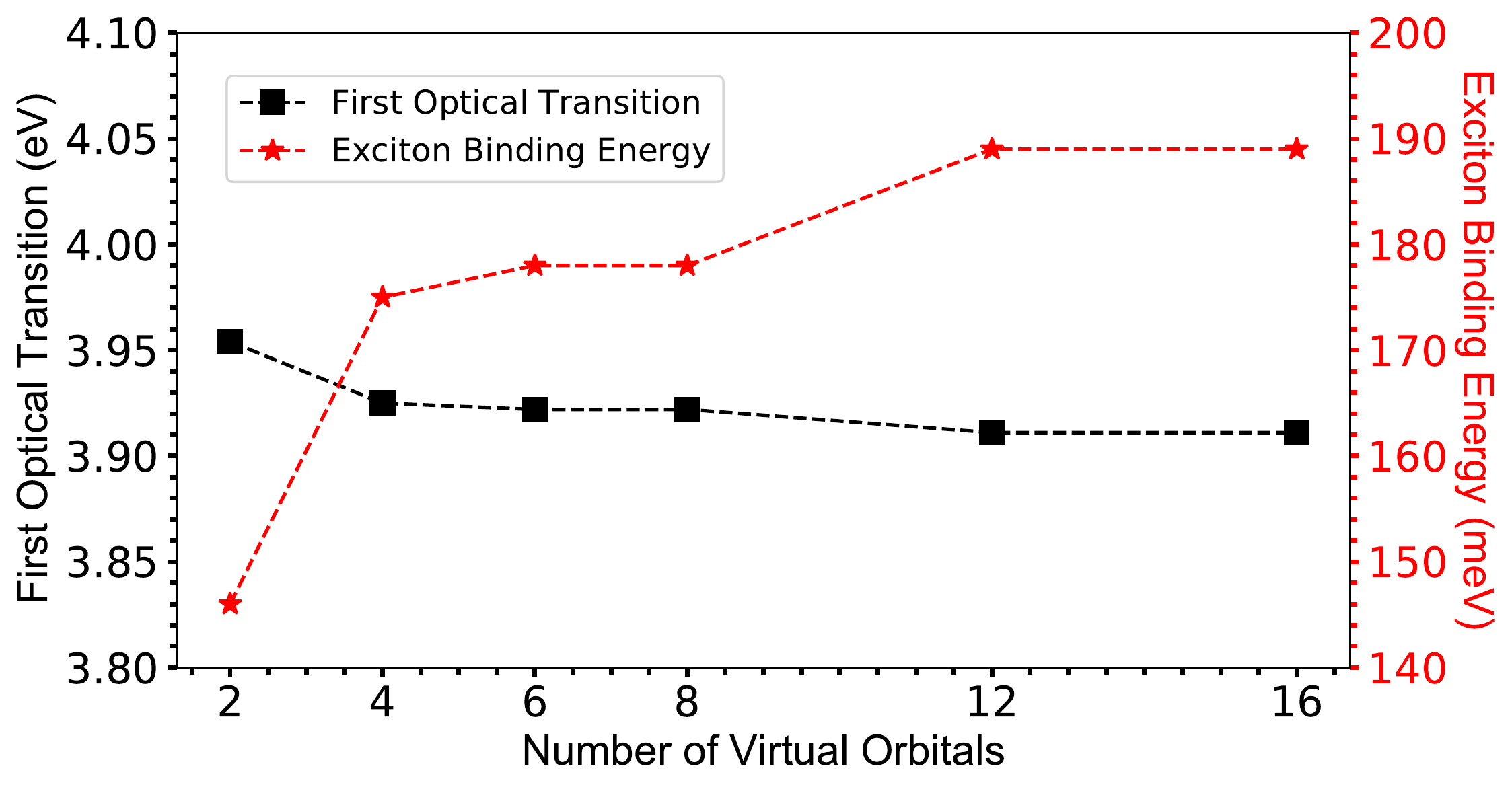}
  \caption{Convergence behaviour of first excitonic transition and exciton binding energy with respect to number of virtual orbitals used in the BSE}
  \label{fig:HSE_Exciton_Binding_Eenergy_Convergence_NBANDSV}
\end{figure}
\begin{figure}
  \centering{}
  \includegraphics[width=0.8\textwidth, keepaspectratio]{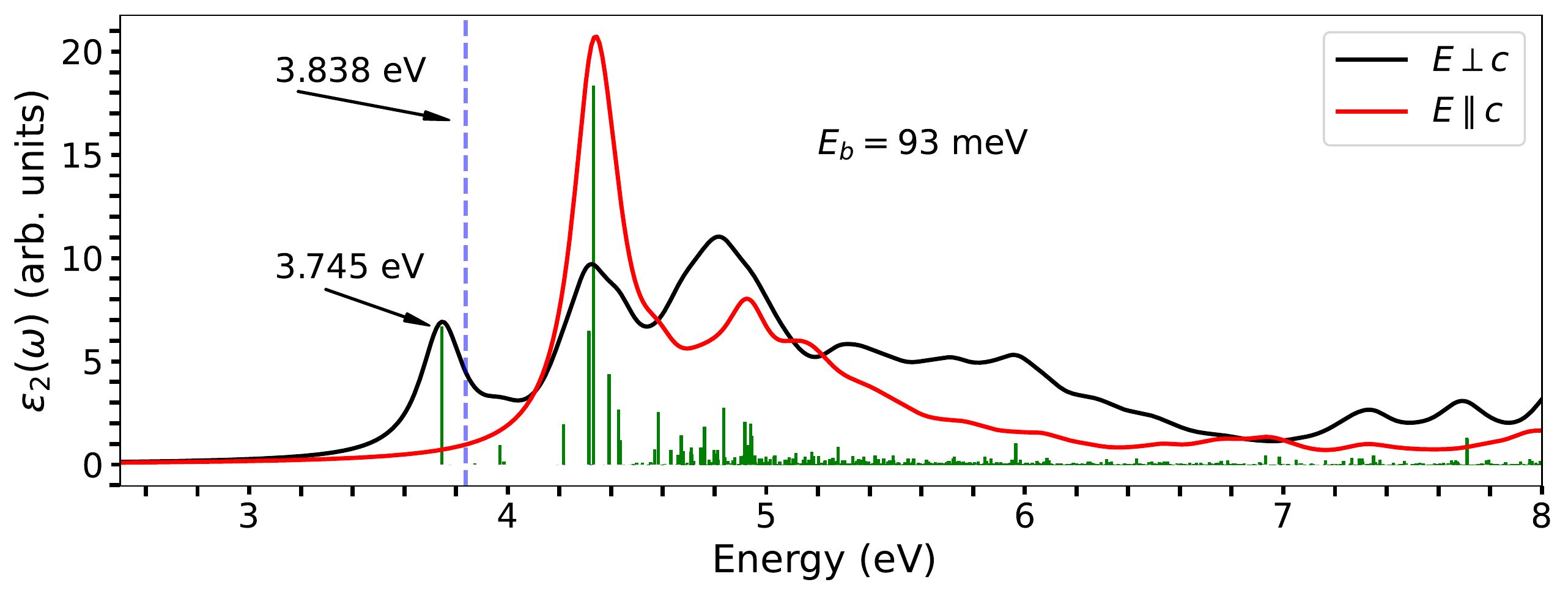}
  \caption{Dielectric function obtained from BSE with PBE starting wavefunction}
  \label{fig:DFT_BSE_Dielectric_DFT_BSE_Oscillator_Strength}
\end{figure}
\begin{figure}
  \centering{}
  \includegraphics[width=0.8\textwidth, keepaspectratio]{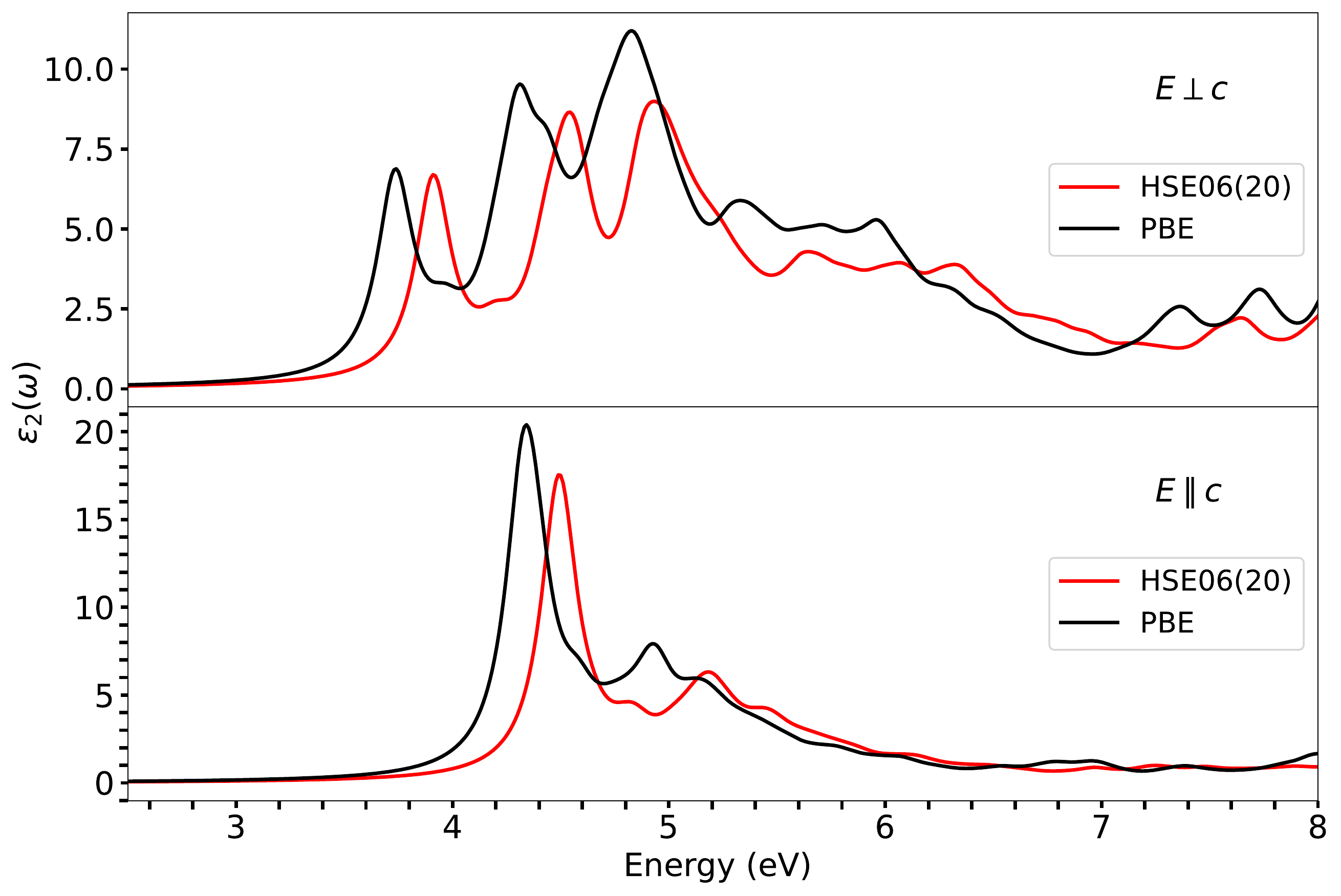}
  \caption{Comparison of dielectric function obtained from BSE with PBE and HSE06(20) starting wavefunction}
  \label{fig:BSE_Spectra_HSE_vs_DFT}
\end{figure}

\begin{figure}
	\centering{}
	\includegraphics[width=0.7\textwidth,keepaspectratio]{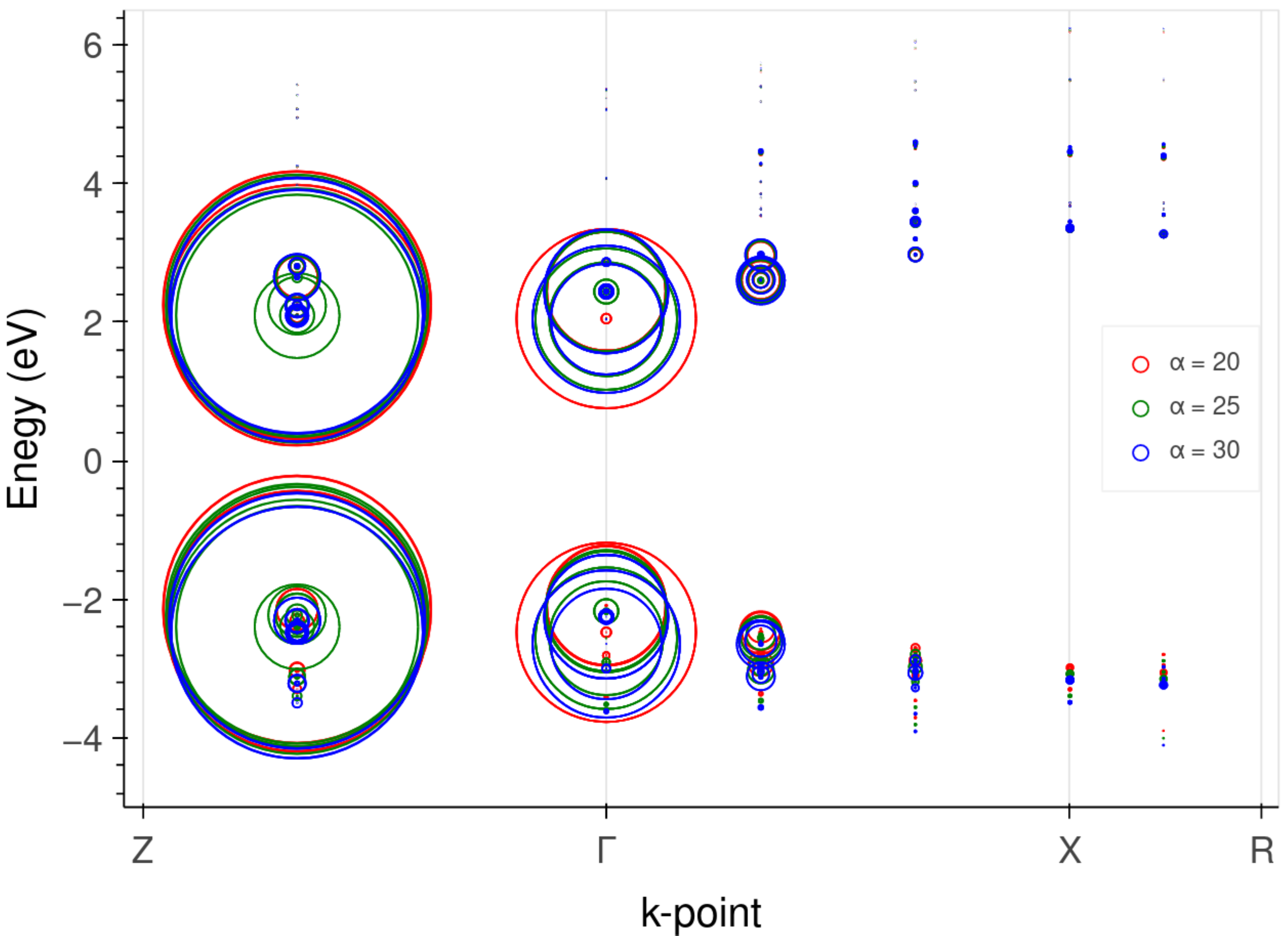}
	\caption{Fatband representation showing region of Brillouin zone that are important for first optical peak in anatase \ce{TiO2} shown for HSE06($\alpha$ = 20, 25, 30\%). Irrespective of the value of $\alpha$, the region of Brillouin zone at which transitions occur and exciton coupling coefficient are similar(except for small change in the energies of individual excitons).}
	\label{fig:Fatband_HSE_AEXX_Combined}
\end{figure}

\begin{figure}
  \centering{}
  \includegraphics[width=0.8\textwidth, keepaspectratio]{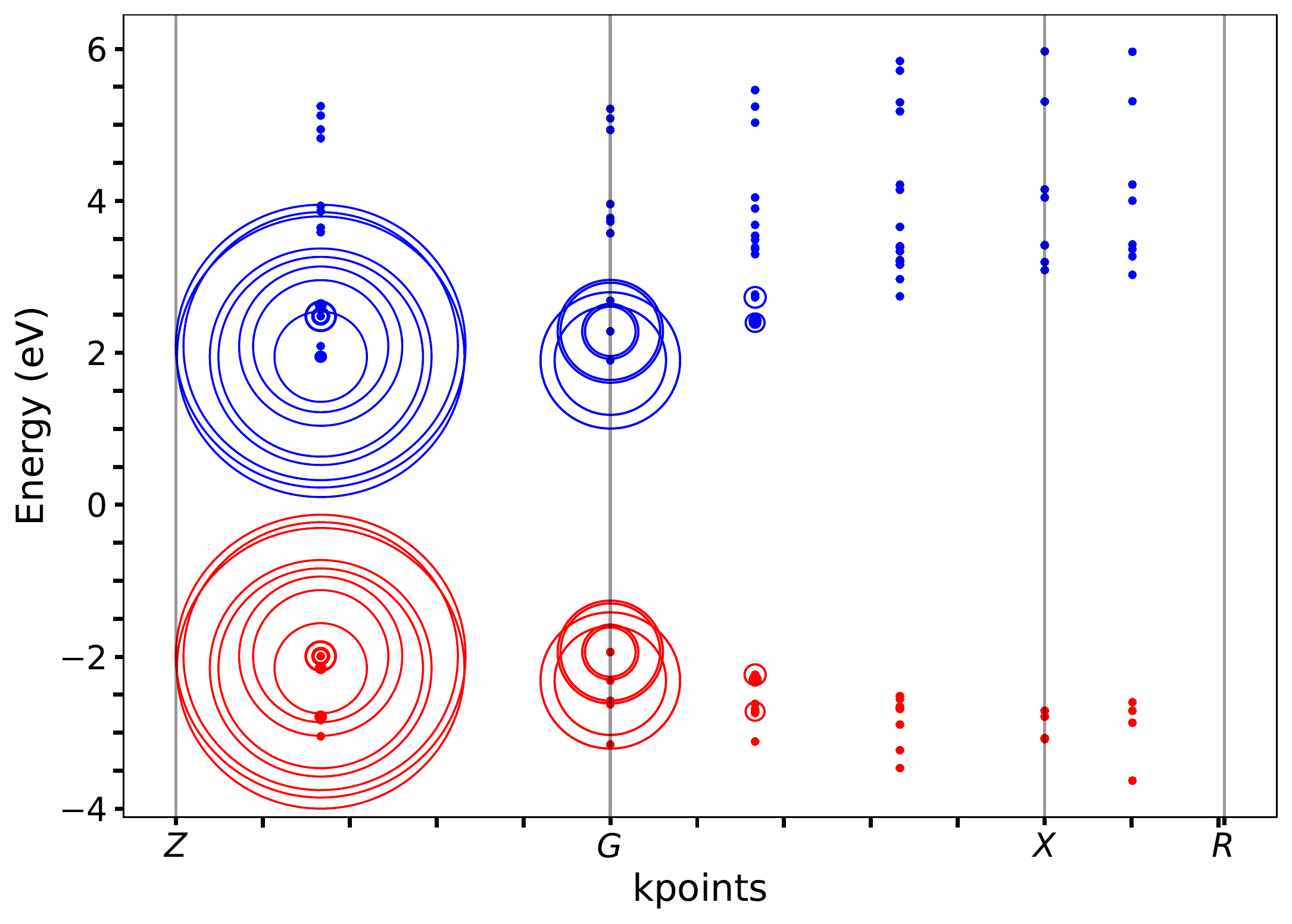}
  \caption{Projection of excitonic amplitudes on the quasiparticle bandstructure near the band edges obtained for PBE starting wavefunction.}
  \label{fig:DFT_BSE_Fatband}
\end{figure}
\end{document}